\documentclass[12pt]{article}
\usepackage{epsfig}
\usepackage{graphicx}
\usepackage{a4}
\usepackage{latexsym}
\usepackage{cite}

\usepackage{color}
\usepackage{colordvi}

\usepackage{pslatex}
\usepackage[latin1]{inputenc}
\usepackage[T1]{fontenc}

\newcommand{\as}{\alpha_{\rm s}}

\def\MSbar{\overline{\mathrm{MS}}}
\def\ep{\epsilon}
\def\nf{{n^{}_{\! f}}}
\def\nl{{n^{}_{\! l}}}
\def\nh{{n^{}_{\! h}}}

\begin{document}
\setlength{\parskip}{0.2cm}
\setlength{\baselineskip}{0.55cm}

\begin{titlepage}
\noindent
\vspace{2.0cm}
\begin{center}
\LARGE {\bf
Inclusive Heavy Flavor Hadroproduction \\
in NLO QCD: the Exact Analytic Result
} \\
\vspace{2.6cm}
\large
M. Czakon$^{a,b}$ and A. Mitov$^{c}$ \\
\vspace{1.4cm}
\normalsize
{\it
$^{a}$Institut f\"ur Theoretische Physik und Astrophysik, Universit\"at
W\"urzburg, \\[0.5ex]
Am Hubland, D-97074 W\"urzburg, Germany \\[.5cm]
$^{b}$Institute of Nuclear Physics, NCSR ``DEMOKRITOS'', \\[0.5ex]
15310 Athens, Greece \\[.5cm]
$^{c}$C. N. Yang Institute for Theoretical Physics, Stony Brook
University, Stony Brook, NY 11794, USA} \vfill
\large {\bf Abstract}
\vspace{-0.2cm}
\end{center}

We present the first exact analytic result for all partonic channels
contributing to the total cross section for the production of a pair
of heavy flavors in hadronic collisions in NLO QCD. Our calculation
is an essential step in the derivation of the top quark pair
production cross section at NNLO in QCD, which is a cornerstone of
the precision LHC program. Our results uncover the analytical
structures behind observables with heavy flavors at higher orders.
They also reveal surprising and non-trivial implications for
kinematics close to partonic threshold.

%
\vspace{3.0cm}
\end{titlepage}

%
%
\section{Introduction}
\label{sec:intro}
Thanks to the Tevatron, the production of heavy flavors at hadron
colliders has nowadays become a very dynamic field of research. The
imminent physics expected from the Large Hadron Collider (LHC)
places even stronger emphasis on this class of processes. The very
large production rates for both top and bottom at the LHC will alow
studies of heavy flavors with precision so far seriously considered
only for processes with light flavors. Accurate knowledge of top
production is fundamental in Higgs and most Beyond the Standard
Model as well as Standard Model studies
\cite{Beneke:2000hk,Kehoe:2007px,Han:2008xb,Bernreuther:2008ju,Laenen:2008jx}.

These very exciting experimental developments have to be confronted
by theory. At present, the next-to-leading order (NLO) corrections in the
strong coupling have been accounted for. That includes the fully
inclusive production rate, the one-particle inclusive cross section,
spin correlations as well as the fully exclusive production and
charge asymmetry
\cite{Nason:1987xz,Beenakker:1988bq,Nason:1989zy,Beenakker:1990maa,Mangano:1991jk,Bernreuther:2001bx,Bernreuther:2004jv,Kuhn:1998jr,Kuhn:1998kw,Almeida:2008ug}.
At the NLO accuracy level several associated production processes
have also been studied
\cite{Beenakker:2001rj,Beenakker:2002nc,Dawson:2002tg,Dawson:2003zu,Baur:2004uw,Baur:2005wi,Dittmaier:2007wz,Dittmaier:2008uj,Lazopoulos:2008de}.

Various sources of uncertainty affect the theoretical predictions
for heavy flavor hadroproduction. The most important one is due to
unknown higher order corrections; these are NNLO effects in the
strong coupling or electroweak effects. Their reduction mandates
higher order calculations.

Partial NNLO corrections to the total cross section have been
presented in the literature
\cite{Kidonakis:2001nj,Moch:2008qy,Kidonakis:2003qe}. They originate
from a fixed order truncation of the all-order exponentiation of
soft gluon logarithms
\cite{Sterman:1986aj,Laenen:1993xr,Catani:1996yz,Bonciani:1998vc,Cacciari:2008zb}.
We would like to emphasize, however, that such an approach to the
fixed order cross section does not represent a controllable
approximation in the sense of being derived from a consistent
expansion in a small parameter. A priori, such partial corrections
cannot be expected to approximate well the exact fixed-order result;
an explicit example can be found in Fig.3 in
Ref.~\cite{Melnikov:2005bx} as well as in Section~\ref{sec:results}
below. A careful investigation of this point is beyond the scope of
the present article. It will be provided elsewhere \cite{CM2}.

In a recent analysis, the authors of Ref.~\cite{Hagiwara:2008df}
point out that bound-state effects may have been underestimated in
past studies of kinematics very close to threshold. They argue that
this effect can have a significant impact on both near-threshold
differential distributions and top mass measurements.

Another source of uncertainty in top production are initial state
non-perturbative effects contained in the parton distribution functions. These
effects have been discussed extensively in the literature; for a comparative
study of the various pdf sets in top quark pair production see for example
Ref.~\cite{Moch:2008qy,Cacciari:2008zb}. The importance of top quark pair
production in constraining pdf sets and other LHC observables has been studied
in  Ref.~\cite{Nadolsky:2008zw}. Overall, pdf-related effects in the
considered process are expected to be in the 3-4 percent range and are smaller
than the missing higher order effects of interest to us in this work.

The present paper represents a needed step in the program
\cite{Czakon:2007wk,Czakon:2007ej,Czakon:2008zk,Korner:2008bn,Kniehl:2008fd,Bonciani:2008az,Anastasiou:2008vd}
for the systematic derivation of the NNLO QCD corrections to the
total heavy flavor production cross section. The phenomenological
importance of the total pair-production cross section was
established in the original papers \cite{Nason:1989zy}, where it was
demonstrated that the effect of the NLO corrections is mainly on the
overall normalizations and not on the {\it shapes} of differential
distributions (however see Ref.~\cite{Hagiwara:2008df} as mentioned
above). It is natural to expect that this feature will persist at
even higher orders, especially for distributions not subjected to
too strong cuts. The reason for this expectation can be traced back
to the non-trivial kinematics of the produced heavy quark pair
starting from the leading order approximation.

Specifically, in this paper we derive the first exact analytic
expression for the NLO QCD corrections to the total inclusive cross
section for heavy quark pair production at hadron colliders. The
results obtained are thus new, since previously only a numerically
extracted fit was available. We comment on the quality of this
numerical approximation in Sec.~\ref{sec:results}. Secondly, we aim
at tuning a calculational approach that is capable of tackling the
derivation of the NNLO corrections for this observable.

The calculation of the top quark pair production cross section at higher
orders presents several significant challenges. First, the analytic structure
of the total cross section is very complicated - and different from the one
encountered so far in multiloop massless calculations. Indeed, radicals start
to show up even at leading order. Our approach in addressing this problem is
discussed in Section~\ref{sec:partonlevel}. Our findings (see
sections.~\ref{sec:results},\ref{sec:singularities},\ref{sec:master}) not only
confirm the expectations, but also demonstrate that a priori unforeseen
analytic structures arise starting from NLO.  Second, at NNLO one also has to
tackle the problem of the large number of three-loop cut diagrams. Third, in
the case of top quark pair production the value of the top mass is not small
compared to the typical partonic center of mass energy. That in turn means
that one has to obtain results, which are exact in the mass. In consequence,
approaches based solely on an expansion in powers of the mass of the heavy
flavor are not likely to be sufficient to cover the whole kinematic range. We
illustrate this point by analyzing the newly derived exact NLO result in
Sec.~\ref{sec:results}.

Finally, we would like to address the question of why we prefer an
analytic approach and do not pursue a purely numerical one from the
very beginning. One reason is that the NLO result, including terms
subleading in the regularization parameter, is a prerequisite for
the NNLO calculation. It is clearly desirable to have complete
control over the lower order result. In fact, as we show in our
following discussion, effects due to uncertainties inherent in
numerical calculations can be sizable enough to require additional
consideration. Our second motivation for pursuing analytical methods
is based on the observation that the cross section we compute
essentially represents a one-variable problem. Therefore, it seems
reasonable to expect that at least most of the calculation should be
feasible in terms of known functions and by using automatic methods.
Our expectations have indeed turned out to be true. As a result we
have gained deep insight into the analytic structure of cross
sections as complex valued functions in the massive case.

The paper is organized as follows: in the next section we introduce some
basic notations and definitions. In Sec.~\ref{sec:partonlevel} we introduce
our calculational method.  Section~\ref{sec:results} contains our main result,
the analytic expressions for all partonic channels contributing at NLO.
Sections~\ref{sec:singularities} and \ref{sec:master}, present detailed
discussions of the analytic features of our results. Finally, in order to be as
self-contained as possible, we collected various known formulae in the
Appendix.

%
%
\section{Setup}
\label{sec:setup}

We calculate the NLO corrections to the total cross section for the
process
\begin{equation}
\label{eq:ppQQ}
p+p({\bar p}) \:\:\rightarrow\:\: Q+ {\bar Q} +X  \, .
\end{equation}
At partonic level, this involves the following reactions
\begin{eqnarray}
  \label{eq:partonicreactions}
q+{\bar q} \:\:\rightarrow\:\: Q+ {\bar Q} +X  \, ,
\nonumber\\
g+g \:\:\rightarrow\:\: Q+ {\bar Q} +X  \, ,
\nonumber\\
g+q \:\:\rightarrow\:\: Q+ {\bar Q} +X  \, .
\end{eqnarray}
Following the established notations, we denote the cross section for
the reaction Eq.~(\ref{eq:ppQQ}) as
\begin{equation}
\label{eq:sigma_tot_hadron} \sigma(S,m^2) = \sum_{ij} \int dx_1 dx_2
\hat\sigma_{ij}(s,m^2,\mu^2) f_i(x_1,\mu^2) f_j(x_2,\mu^2) \, ,
\end{equation}
where $s=x_1x_2S$ is the partonic center of mass energy, $S$ is the
corresponding hadronic invariant and $m$ is the pole mass of the
heavy quark. The physical region is defined by $s\geq 4m^2 > 0$. The
scale $\mu$ denotes both the renormalization and factorization
scales as appropriate. For simplicity, we do not distinguish between
the two. The functions $f_i$ are the parton distribution functions
for parton $i$ in the (anti)proton.

The partonic cross section $\hat\sigma$ can be written as an
expansion in the strong coupling. We will keep the traditional
notation, where the cross sections are written in terms of the
following dimensionless functions
\begin{equation}
\label{eq:sigma_hat} \hat\sigma_{ij}(s,m^2,\mu^2) = {\as^2(\mu^2)
\over m^2} \, f_{ij}\left({m^2\over s},{\mu^2\over m^2}\right)\, ,
\end{equation}
where
\begin{equation}
\label{eq:f_ij} f_{ij}\left(\frac{m^2}{s}, \frac{\mu^2}{m^2}\right)
= f^{(0)}_{ij}\left(\frac{m^2}{s}\right) + 4 \pi \alpha_s(\mu^2)
\left( f^{(1)}_{ij}\left(\frac{m^2}{s}\right) +
\bar{f}^{(1)}_{ij}\left(\frac{m^2}{s}\right) \log
\left(\frac{\mu^2}{m^2}\right) \right) + {\cal O} \left(\alpha_s^2
\right) \; .
\end{equation}

The partonic cross section $\hat\sigma$ contains three implicit
scheme dependencies. First, it depends on the definition of the
strong coupling $\as$. We work in terms of the standard $\MSbar$
coupling defined through
\begin{eqnarray}
\label{eq:alpha-s-renorm}
\as^{\rm{b}} S_\epsilon \: = \: \as
\biggl[
   1
   - {\beta_0 \over \epsilon} \biggl( {\as \over 2 \pi} \biggr)
   + {\cal O}(\as^2)
  \biggr]
\, ,
\end{eqnarray}
where $S_\epsilon=(4 \pi)^\ep \exp(-\ep \gamma_{\rm E})$ and $\beta_0 = {11
  \over 6}\*C_A - {2 \over 3}\*T_F\*\nf$ is the first term of the QCD
$\beta$-function (known by now up to the four-loop level
\cite{vanRitbergen:1997va,Czakon:2004bu}). The color factors in an
${\rm{SU}}(N)$-gauge theory are $C_A = N$, $C_F = (N^2-1)/(2\*N)$
and $T_F = 1/2$. Throughout this paper $\nf$ stands for the total
number of flavors, which is the sum of $\nl$ light and $\nh=1$ heavy
quarks. Switching between definitions of the coupling with $\nl+1$
(as adopted in this paper) and $\nl$ active flavors is done with the
standard decoupling relations (also known up to the four-loop level
\cite{Schroder:2005hy,Chetyrkin:2005ia}, see Appendix \ref{sec:appA} for the
expression needed in our case).

Second, the partonic cross section $\hat\sigma$ depends on the
definition of the mass $m$. As we mentioned above, we work in terms
of the usual pole mass \cite{Melnikov:2000qh,Chetyrkin:1999qi}. The
third implicit dependence is on the scheme used to factor out the
collinear (mass) singularities related to initial state collinear
radiation. We adopt the standard $\MSbar$ factorization scheme
defined through
\begin{equation}
\label{eq:mass-factorization} {\hat\sigma_{ij}(\rho)\over \rho} =
\sum_{kl} \left[ \left({\sigma_{kl}(\tau,\ep) \over \tau}\right) \,
\otimes \, \Gamma_{ki}(\tau,\ep) \, \otimes \, \Gamma_{lj}(\tau,\ep)
\right] (\rho) \, ,
\end{equation}
where $\rho = 4m^2/s$ with $0<\rho\leq 1$. The variable $\tau, \,
\rho \leq \tau \leq 1,$ stands for the corresponding (dummy)
convolution variable. The functions $\sigma_{kl}$ are the ``bare"
(i.e. collinearly un-renormalized) hard scattering cross sections
and the collinear counterterms $\Gamma$ are given by
\begin{equation}
\label{eq:coll-counterterm} \Gamma_{ij}(\tau,\ep) = \delta_{ij}
\delta(1-\tau) - \left({\as^{(\nl)}(\mu^2)\over 2\pi}\right)
{P^{(0,\nl)}_{ij}(\tau)\over \ep} + {\cal O}(\as^2) \, ,
\end{equation}
where $P^{(n,n_l)}(\tau)$ are the $(n+1)$-loop (space-like) QCD
splitting functions
\cite{Altarelli:1977zs,Curci:1980uw,Furmanski:1980cm,Floratos:1981hs,Moch:2004pa,Vogt:2004mw,Vermaseren:2005qc,Mitov:2006ic}.
The superscript $\nl$ in $P^{(n)}$ and $\as$ emphasizes that these
quantities are evaluated in a scheme with $\nl$-flavors.

%
%
\section{Parton level evaluation}
\label{sec:partonlevel}

In this section we describe the evaluation of the partonic
cross section $\sigma_{ij}(\rho,\ep)$ appearing in
Eq.~(\ref{eq:mass-factorization}). Following the method introduced
in \cite{Anastasiou:2002yz} we perform simultaneously all loop and
real radiation integrations. With the help of the IBP identities
\cite{Chetyrkin:1981qh,Tkachov:1981wb} which we solve with the
Laporta algorithm \cite{Laporta:2001dd}, the whole problem is mapped
into a total of 37 master integrals.

The needed partonic cross sections are non-trivial functions of a
single dimensionless variable $\rho=4m^2/s$
\footnote{The dependence on the renormalization/factorization scale
$\mu^2/m^2$ requires only calculations of one order lower and are
thus considered 'trivial'. However, see the discussion at the end of
this section.}
. As was established in the literature, the analytic structure of
the result becomes rather complicated. In particular, even at LO
radicals like $\beta = \sqrt{1-\rho}$ appear which suggests that the
application of established analytic methods in this problem may be
too cumbersome. To rectify the situation we perform the following
change of variables
\begin{eqnarray}
\label{eq:x-deff} {m^2\over s} &=& {x\over (1+x)^2} \,\,
,\nonumber\\
x &=& { 1-\sqrt{1-{4m^2\over s}} \over 1+\sqrt{1-{4m^2\over s}} }
\,\, .
\end{eqnarray}

This change of variables allows one to express the results for the
partonic cross sections almost entirely in terms of harmonic
polylogarithms (HPL) \cite{Remiddi:1999ew} of the variable $x$ as
well as rational functions of $x$ and $1\pm x$. Such analytic
structures have been widely explored and can be dealt with in a
straightforward manner. A particularly welcoming feature of this
approach is that one can achieve a large degree of automatization in
the solving of the differential equations for the masters with the
help of existing software \cite{Maitre:2005uu,Maitre:2007kp}.

The reason why this change of variables is helpful can be seen as follows. The
physical cross section for the production of two heavy particles in the final
state has two physical singularities: the points $m^2=0$ representing the high
energy (or massless) limit, and the kinematic threshold for heavy pair
production $s=4m^2$. The third singularity, which is not in the physical
region, is $\vert m^2/s \vert \to \infty$. The transformation
(\ref{eq:x-deff}) maps these three singularities into the points $x=(0,1,-1)$
which are precisely the singular points of the differential equations defining
the set of HPL's.

As we will see in the following, however, the differential equations
for a few of the master integrals posses singularities other than the
three described above. These are the points $s=m^2, s=-m^2, s=-4m^2$
and $s=-16m^2$. They are all outside the physical region $0 <
4m^2/s \leq 1$. To the best of our knowledge, this is the first
calculation of an observable with massive flavors where such
pseudo-threshold-like singular points have been encountered. At NLO
they appear only in the $g+g\to Q{\overline Q}+X$ subprocess
starting from order {\cal O($\epsilon^0$)}. We discuss them in
detail in Sec.\ref{sec:singularities}.

To completely fix the solutions for the master integrals one has to
supply suitable boundary conditions. The behavior of the solution
around any one of the singularities $x=(0,\pm 1)$ can be used. In
the cases where integrals form a system of two or three differential
equations it turns out that one needs to supply information about
the behavior of the solution at more than one singular point. It is
also worth mentioning that the approach used in
Ref.~\cite{Bonciani:2003te,Bonciani:2003hc} to fix the boundary
conditions by simply requiring regularity of the solution cannot be
applied to the problem at hand since the masters are typically
singular at all singular points. A detailed discussion of the most
complicated master we encountered in this calculation is given in
Sec.~\ref{sec:master}. Here, let us only note that while some of the
boundaries required a direct calculation by Mellin-Barnes methods
\cite{Smirnov:1999gc,Tausk:1999vh}, where we used the {\tt MB}
package \cite{Czakon:2005rk}, many could be simply obtained by
requiring the vanishing of the integral at threshold.

The evaluation of the convolutions in
Eq.~(\ref{eq:mass-factorization}) also deserves a comment. A
complication arises from the fact that while the dummy variable $\tau$
is the 'natural' variable for the splitting functions entering the
collinear counterterms $\Gamma$, the same variable is {\it not} the
natural one for the partonic cross sections $\sigma$. To write the
latter in their natural variables one has to switch from the variable
$\rho$ to the conformal variable $x$ defined in
Eq.~(\ref{eq:x-deff}).

%
%
\section{Results}
\label{sec:results}

At this point, we are ready to present our main result, namely the
analytic expressions for the partonic cross sections at
next-to-leading order. Our calculation has been performed with full
account of the dependence on the number of colors $N$. Because of
the substantial length of the formulae, however, we will present
here only the result for $N=3$. The complete results, valid for any
$SU(N)$ color gauge group are available in electronic form in
\cite{MathematicaFile}.
{\small
\begin{eqnarray}
&& \!\!\!\! \!\!\!\! \!\!\!\! \!\!\!\! \!\!\!\! \!\!\!
  \label{eq:fijqq}
  4 \pi f^{(1)}_{q\bar{q}}(x) = \nonumber \\ &&
  \left(-\frac{1061}{243 (x+1)}+\frac{193}{81 (x+1)^2}+\frac{8486}{243
    (x+1)^3}-\frac{4660}{81 (x+1)^4}+\frac{2312}{81 (x+1)^5}-\frac{448}{81
    (x+1)^6}+\frac{128}{81 (x+1)^7}\right) \nonumber \\
  &+& \left(\frac{1046}{243 (x+1)}-\frac{830}{243 (x+1)^2}-\frac{2380}{81
    (x+1)^3}+\frac{11540}{243 (x+1)^4}-\frac{4616}{243 (x+1)^5}\right){\rm H}(-1,x)
  \nonumber \\
  &+& \left(-\frac{2834}{243 (x+1)}+\frac{1346}{243 (x+1)^2}+\frac{4918}{81
    (x+1)^3}-\frac{2470}{27 (x+1)^4}+\frac{2924}{81 (x+1)^5}+\frac{448}{81
    (x+1)^6}-\frac{512}{81 (x+1)^7} \right. \nonumber \\ &&
  \left. +\frac{128}{81 (x+1)^8}\right){\rm H}(0,x)
  \nonumber \\
  &+& \left(-\frac{2344}{243 (x+1)}+\frac{1912}{243 (x+1)^2}+\frac{5264}{81
    (x+1)^3}-\frac{25600}{243 (x+1)^4}+\frac{10240}{243 (x+1)^5}\right){\rm H}(1,x)
  \nonumber \\
  &+& \left(\frac{100}{243 (x+1)}-\frac{208}{243 (x+1)^2}-\frac{118}{81
    (x+1)^3}+\frac{914}{243 (x+1)^4}-\frac{506}{243 (x+1)^5}+\frac{2}{9
    (x+1)^6}\right)\pi ^2 \nonumber \\
  &+& \left(\frac{200}{81 (x+1)}+\frac{232}{81 (x+1)^2}-\frac{1600}{81
    (x+1)^3}+\frac{1616}{81 (x+1)^4}-\frac{416}{81 (x+1)^5}-\frac{32}{81
    (x+1)^6}\right){\rm H}(-1,0,x) \nonumber \\
  &+& \left(\frac{256}{81 (x+1)}-\frac{256}{81 (x+1)^2}-\frac{512}{27
    (x+1)^3}+\frac{2560}{81 (x+1)^4}-\frac{1024}{81 (x+1)^5}\right){\rm H}(-1,1,x)
  \nonumber \\
  &+& \left(\frac{128}{81 (x+1)}-\frac{128}{81 (x+1)^2}-\frac{584}{81
    (x+1)^3}+\frac{728}{81 (x+1)^4}+\frac{40}{81 (x+1)^5}-\frac{184}{81
    (x+1)^6}\right){\rm H}(0,-1,x) \nonumber \\
  &+& \left(-\frac{40}{27 (x+1)}+\frac{40}{27 (x+1)^2}+\frac{148}{27
    (x+1)^3}-\frac{596}{81 (x+1)^4}+\frac{44}{81 (x+1)^5}+\frac{4}{3
    (x+1)^6}\right){\rm H}(0,0,x) \nonumber \\
  &+& \left(-\frac{128}{27 (x+1)}+\frac{272}{27 (x+1)^2}+\frac{1504}{81
    (x+1)^3}-\frac{400}{9 (x+1)^4}+\frac{64}{3 (x+1)^5}-\frac{64}{81
    (x+1)^6}\right){\rm H}(0,1,x) \nonumber \\
  &+& \left(\frac{256}{81 (x+1)}-\frac{256}{81 (x+1)^2}-\frac{512}{27
    (x+1)^3}+\frac{2560}{81 (x+1)^4}-\frac{1024}{81 (x+1)^5}\right){\rm H}(1,-1,x)
  \nonumber \\
  &+& \left(-\frac{368}{81 (x+1)}+\frac{368}{81 (x+1)^2}+\frac{2240}{81
    (x+1)^3}-\frac{1216}{27 (x+1)^4}+\frac{448}{27 (x+1)^5}+\frac{64}{81
    (x+1)^6}\right){\rm H}(1,0,x) \nonumber \\
  &+& \left(-\frac{512}{81 (x+1)}+\frac{512}{81 (x+1)^2}+\frac{1024}{27
    (x+1)^3}-\frac{5120}{81 (x+1)^4}+\frac{2048}{81
    (x+1)^5}\right){\rm H}(1,1,x) \nonumber\\
%
%
&+& \nl~\left\{ - \frac{8 x \left(x^3+3 x^2-3 x-1\right) \left(6
{\rm H}(-1,x)-3 {\rm H}(0,x)-5\right)}{243 (x+1)^5} \right\} \; ,
\end{eqnarray}
}

\begin{figure}[t]
  \begin{center}
    \epsfig{file=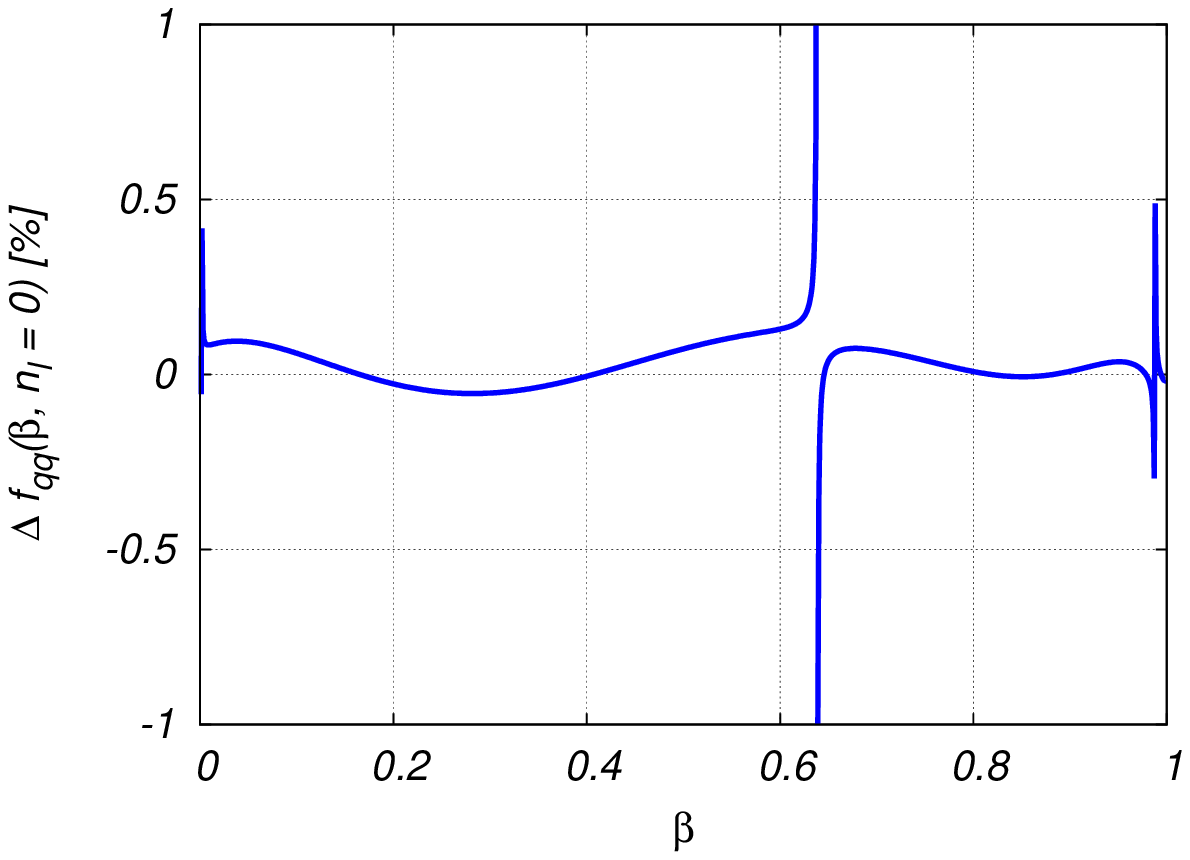,width=8cm}
    \epsfig{file=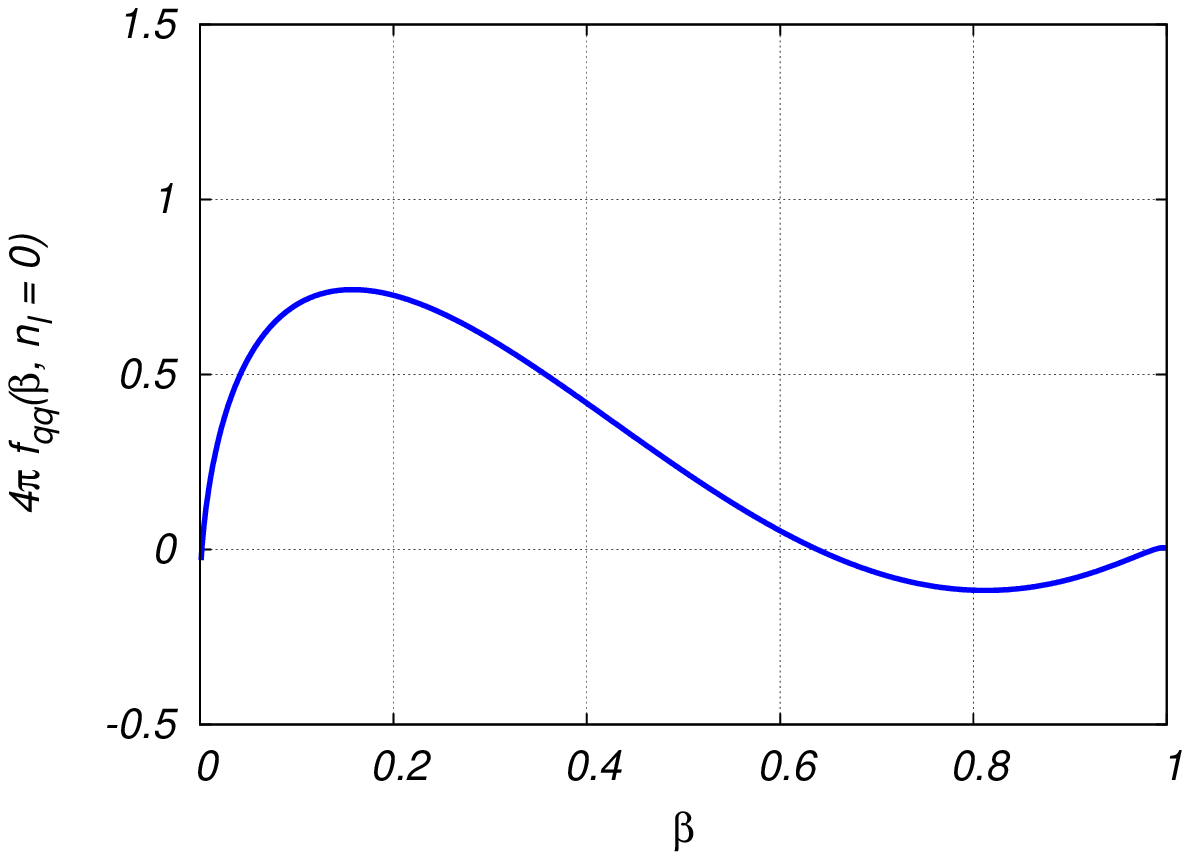,width=8cm}
  \end{center}
  \caption{\it \label{fig:qq} Error implied by the numerical integration of the
    cross section \cite{Nason:1987xz} in the quark-anti-quark production channel
    against the exact result. The functions are defined in
    Eqns.~(\ref{eq:fijqq},\ref{eq:error}).}
\end{figure}

{\small
\begin{eqnarray}
  && \!\!\!\! \!\!\!\! \!\!\!\! \!\!\!\! \!\!\!\! \!\!\!
    \label{eq:fijgq}
  4 \pi f^{(1)}_{gq}(x) = \nonumber \\ &&
  \left(-\frac{7291}{24300}+\frac{167}{405 (x+1)}+\frac{112}{81
    (x+1)^2}-\frac{4439}{1215 (x+1)^3}+\frac{3319}{810
    (x+1)^4}-\frac{3319}{2025 (x+1)^5}\right) \nonumber \\
  &+& \left(-\frac{181}{810}+\frac{383}{81 (x+1)}-\frac{1568}{81
    (x+1)^2}+\frac{2804}{81 (x+1)^3}-\frac{2638}{81 (x+1)^4}+\frac{5276}{405
    (x+1)^5}\right){\rm H}(-1,x) \nonumber \\
  &+& \left(\frac{181}{270}-\frac{974}{81 (x+1)}+\frac{7321}{162
    (x+1)^2}-\frac{2173}{27 (x+1)^3}+\frac{2117}{27 (x+1)^4}-\frac{5113}{135
    (x+1)^5}+\frac{55}{9 (x+1)^6}\right){\rm H}(0,x) \nonumber \\
  &+& \left(\frac{181}{405}-\frac{766}{81 (x+1)}+\frac{3136}{81
    (x+1)^2}-\frac{5608}{81 (x+1)^3}+\frac{5276}{81 (x+1)^4}-\frac{10552}{405
    (x+1)^5}\right){\rm H}(1,x) \nonumber \\
  &+& \left(-\frac{2}{81 (x+1)}-\frac{7}{324 (x+1)^2}+\frac{25}{54
    (x+1)^3}-\frac{125}{108 (x+1)^4}+\frac{10}{9 (x+1)^5}-\frac{10}{27
    (x+1)^6}\right)\pi ^2 \nonumber \\
  &+& \left(\frac{94}{27 (x+1)}-\frac{190}{27 (x+1)^2}+\frac{16}{3
    (x+1)^3}+\frac{16}{9 (x+1)^4}-\frac{16}{3 (x+1)^5}+\frac{16}{9
    (x+1)^6}\right){\rm H}(-1,0,x) \nonumber \\
  &+& \left(-\frac{34}{27 (x+1)}+\frac{61}{27 (x+1)^2}+\frac{2}{27
    (x+1)^3}-\frac{47}{9 (x+1)^4}+\frac{56}{9 (x+1)^5}-\frac{56}{27
    (x+1)^6}\right){\rm H}(0,-1,x) \nonumber \\
  &+& \left(\frac{4}{27 (x+1)}+\frac{7}{54 (x+1)^2}-\frac{25}{9
    (x+1)^3}+\frac{125}{18 (x+1)^4}-\frac{20}{3 (x+1)^5}+\frac{20}{9
    (x+1)^6}\right){\rm H}(0,0,x) \nonumber \\
  &+& \left(\frac{68}{27 (x+1)}-\frac{122}{27 (x+1)^2}-\frac{4}{27
    (x+1)^3}+\frac{94}{9 (x+1)^4}-\frac{112}{9 (x+1)^5}+\frac{112}{27
    (x+1)^6}\right){\rm H}(0,1,x) \nonumber \\
  &+& \left(\frac{8}{27 (x+1)}-\frac{8}{9 (x+1)^2}+\frac{32}{27
    (x+1)^3}-\frac{16}{27 (x+1)^4}\right)\pi ^2 {\rm H}(-1,x) \nonumber \\
  &+& \left(-\frac{4}{27 (x+1)}+\frac{4}{9 (x+1)^2}-\frac{16}{27
    (x+1)^3}+\frac{8}{27 (x+1)^4}\right)\pi ^2 {\rm H}(0,x) \nonumber \\
  &+& \left(-\frac{16}{9 (x+1)}+\frac{16}{3 (x+1)^2}-\frac{64}{9
    (x+1)^3}+\frac{32}{9 (x+1)^4}\right){\rm H}(-1,-1,0,x) \nonumber \\
  &+& \left(\frac{16}{9 (x+1)}-\frac{16}{3 (x+1)^2}+\frac{64}{9
    (x+1)^3}-\frac{32}{9 (x+1)^4}\right){\rm H}(-1,0,-1,x) \nonumber \\
  &+& \left(-\frac{16}{9 (x+1)}+\frac{16}{3 (x+1)^2}-\frac{64}{9
    (x+1)^3}+\frac{32}{9 (x+1)^4}\right){\rm H}(-1,0,0,x) \nonumber \\
  &+& \left(-\frac{32}{9 (x+1)}+\frac{32}{3 (x+1)^2}-\frac{128}{9
    (x+1)^3}+\frac{64}{9 (x+1)^4}\right){\rm H}(-1,0,1,x) \nonumber \\
  &+& \left(\frac{8}{9 (x+1)}-\frac{8}{3 (x+1)^2}+\frac{32}{9
    (x+1)^3}-\frac{16}{9 (x+1)^4}\right){\rm H}(0,-1,0,x) \nonumber \\
  &+& \left(-\frac{8}{9 (x+1)}+\frac{8}{3 (x+1)^2}-\frac{32}{9
    (x+1)^3}+\frac{16}{9 (x+1)^4}\right){\rm H}(0,0,-1,x) \nonumber \\
  &+& \left(\frac{8}{9 (x+1)}-\frac{8}{3 (x+1)^2}+\frac{32}{9
    (x+1)^3}-\frac{16}{9 (x+1)^4}\right){\rm H}(0,0,0,x) \nonumber \\
  &+& \left(\frac{16}{9 (x+1)}-\frac{16}{3 (x+1)^2}+\frac{64}{9
    (x+1)^3}-\frac{32}{9 (x+1)^4}\right){\rm H}(0,0,1,x) \; ,
\end{eqnarray}
}

\begin{figure}[t]
  \begin{center}
    \epsfig{file=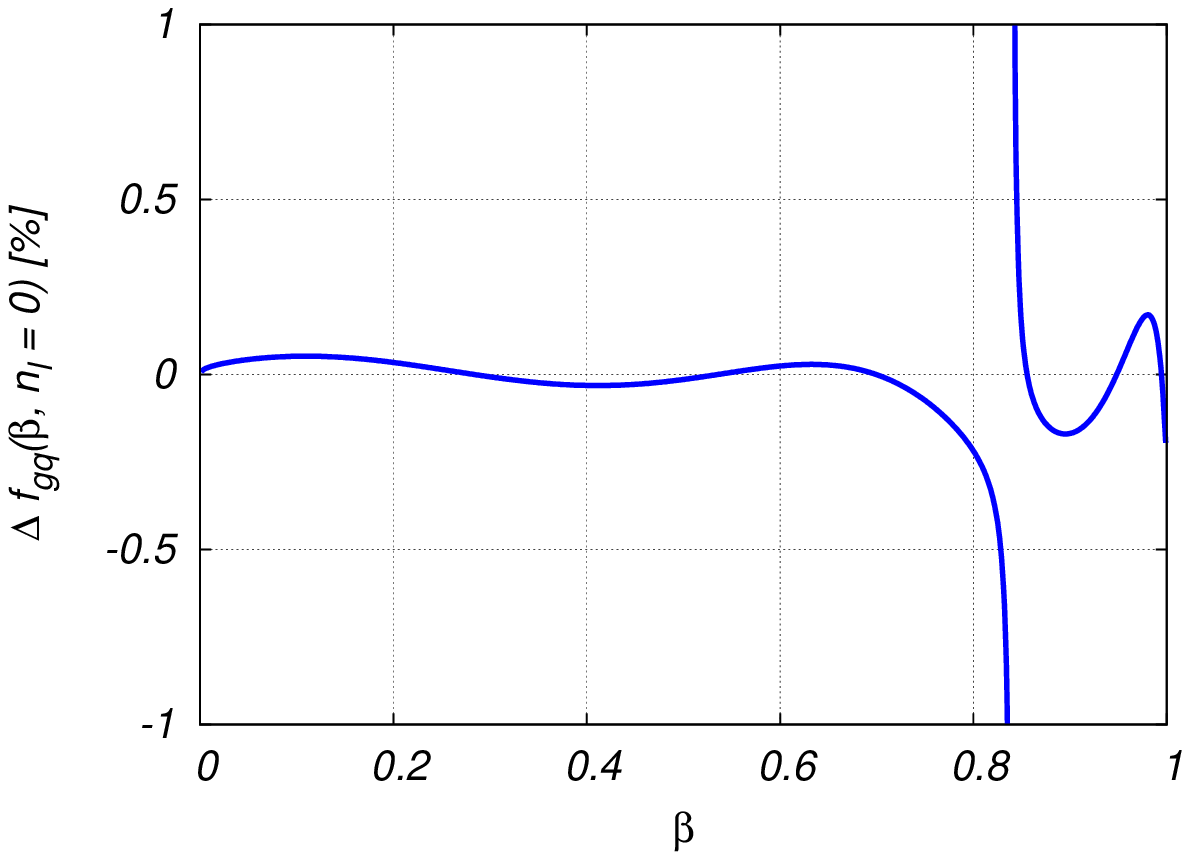,width=8cm}
    \epsfig{file=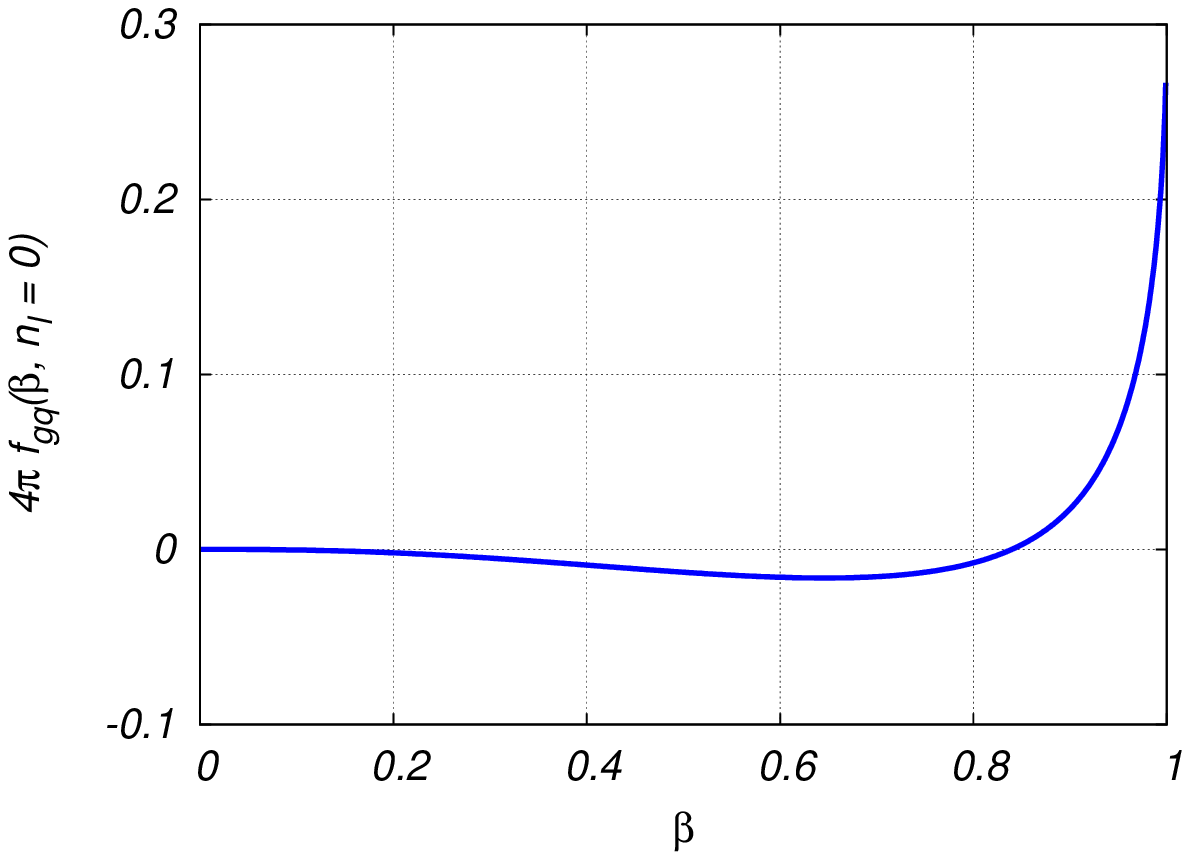,width=8cm}
  \end{center}
  \caption{\it \label{fig:gq} Error implied by the numerical integration of the
    cross section \cite{Nason:1987xz} in the quark-gluon production channel
    against the exact result. The functions are defined in
    Eqns.~(\ref{eq:fijgq},\ref{eq:error}).}
\end{figure}

{\small
\begin{eqnarray}
  && \!\!\!\! \!\!\!\! \!\!\!\! \!\!\!\! \!\!\!\! \!\!\!
  \label{eq:fijgg}
  4 \pi f^{(1)}_{gg}(x) = \nonumber \\ &&
  \left(\frac{13121}{1920 (x+1)}+\frac{38761}{1152 (x+1)^2}-\frac{1506193}{8640
    (x+1)^3}+\frac{312977}{1440 (x+1)^4}-\frac{237377}{3600
    (x+1)^5}-\frac{21}{(x+1)^6} \right. \nonumber \\ &&
  \left. +\frac{6}{(x+1)^7}-\frac{7291}{5400}-\frac{9}{64
    (x-1)}\right) \nonumber \\
  &+& \left(\frac{x+2}{18 \left(x^2+x+1\right)}-\frac{9}{32 (x-1)}+\frac{11227}{576
    (x+1)}-\frac{81}{32 (x-1)^2}-\frac{13999}{64 (x+1)^2}-\frac{27}{16
    (x-1)^3}+\frac{202939}{288 (x+1)^3} \right. \nonumber \\ &&
  \left. -\frac{80471}{96
    (x+1)^4}+\frac{80471}{240 (x+1)^5}-\frac{181}{180}\right){\rm H}(-1,x)
  \nonumber \\
  &+& \left(\frac{x-1}{18 \left(x^2+x+1\right)}-\frac{24665}{576 (x+1)}+\frac{5
    x}{18 \left(x^2+3 x+1\right)}+\frac{301319}{576 (x+1)^2}-\frac{881899}{576
    (x+1)^3}+\frac{1009037}{576 (x+1)^4} \right. \nonumber \\ &&
  \left. -\frac{1173109}{1440
    (x+1)^5}+\frac{2311}{18
    (x+1)^6}-\frac{24}{(x+1)^7}+\frac{6}{(x+1)^8}+\frac{181}{60}+\frac{9}{8
    (x-1)}+\frac{9}{4 (x-1)^2} \right. \nonumber \\ && \left. +\frac{27}{32
    (x-1)^3}\right){\rm H}(0,x) \nonumber \\
  &+& \left(\frac{181}{90}-\frac{260}{9 (x+1)}+\frac{3784}{9
    (x+1)^2}-\frac{12898}{9 (x+1)^3}+\frac{15563}{9 (x+1)^4}-\frac{31126}{45
    (x+1)^5}\right){\rm H}(1,x) \nonumber \\
  &+& \left(-\frac{5 (3 x+1)}{108 \left(x^2+3 x+1\right)^2}+\frac{27}{128
    (x-1)}-\frac{10165}{3456 (x+1)}+\frac{5 (11 x+2)}{216 \left(x^2+3
    x+1\right)}-\frac{135}{128 (x-1)^2} \right. \nonumber \\ &&
  \left. +\frac{26279}{3456 (x+1)^2}-\frac{81}{32
    (x-1)^3}+\frac{17479}{1728 (x+1)^3}-\frac{81}{64 (x-1)^4}-\frac{34091}{864
    (x+1)^4}+\frac{5999}{192 (x+1)^5} \right. \nonumber \\ &&
  \left. -\frac{10049}{1728 (x+1)^6}-\frac{3}{4
    (x+1)^7}\right)\pi ^2 \nonumber \\
  &+& \left(\frac{x+1}{6 \left(x^2+x+1\right)^2}+\frac{63}{32 (x-1)}+\frac{11
    x-2}{12 \left(x^2+x+1\right)}+\frac{5 (11 x+2)}{18 \left(x^2+3
    x+1\right)}-\frac{315}{32 (x-1)^2}-\frac{5 (3 x+1)}{9 \left(x^2+3
    x+1\right)^2} \right. \nonumber \\ && \left. -\frac{189}{8
    (x-1)^3}-\frac{189}{16 (x-1)^4}-\frac{607}{288
    (x+1)}-\frac{14245}{288 (x+1)^2}+\frac{5263}{36 (x+1)^3}-\frac{6533}{72
    (x+1)^4}-\frac{221}{6 (x+1)^5} \right. \nonumber \\ &&
  \left. +\frac{593}{18 (x+1)^6}\right){\rm H}(-1,0,x) \nonumber \\
  &+& \left(-\frac{14}{x+1}-\frac{20}{(x+1)^2}+\frac{220}{(x+1)^3}
  -\frac{310}{(x+1)^4}+\frac{124}{(x+1)^5}\right){\rm H}(-1,1,x)
  \nonumber \\
  &+& \left(\frac{x+1}{6 \left(x^2+x+1\right)^2}-\frac{9}{16 (x-1)}+\frac{11
    x-2}{12 \left(x^2+x+1\right)}+\frac{45}{16 (x-1)^2}+\frac{27}{4
    (x-1)^3}+\frac{27}{8 (x-1)^4}-\frac{833}{48 (x+1)} \right. \nonumber \\ &&
  \left. +\frac{2429}{72
    (x+1)^2}+\frac{11425}{144 (x+1)^3}-\frac{16555}{72 (x+1)^4}+\frac{2725}{16
    (x+1)^5}-\frac{1733}{48 (x+1)^6}\right){\rm H}(0,-1,x) \nonumber \\
  &+& \left(\frac{2-11 x}{12 \left(x^2+x+1\right)}-\frac{45}{64
    (x-1)}+\frac{10769}{576 (x+1)}-\frac{5 (11 x+2)}{36 \left(x^2+3
    x+1\right)}+\frac{225}{64 (x-1)^2}-\frac{23599}{576 (x+1)^2}
  \right. \nonumber \\ && \left. +\frac{-x-1}{6
    \left(x^2+x+1\right)^2}+\frac{5 (3 x+1)}{18 \left(x^2+3
    x+1\right)^2}+\frac{135}{16 (x-1)^3}-\frac{19513}{288 (x+1)^3}+\frac{135}{32
    (x-1)^4}+\frac{32605}{144 (x+1)^4} \right. \nonumber \\ &&
  \left. -\frac{16219}{96 (x+1)^5}+\frac{2693}{96
    (x+1)^6}+\frac{9}{2 (x+1)^7}\right){\rm H}(0,0,x) \nonumber \\
  &+& \left(\frac{361}{9 (x+1)}-\frac{1193}{18 (x+1)^2}-\frac{2354}{9
    (x+1)^3}+\frac{1291}{2 (x+1)^4}-\frac{1331}{3 (x+1)^5}+\frac{773}{9
    (x+1)^6}\right){\rm H}(0,1,x) \nonumber \\
  &+& \left(-\frac{14}{x+1}-\frac{20}{(x+1)^2}+\frac{220}{(x+1)^3}
  -\frac{310}{(x+1)^4}+\frac{124}{(x+1)^5}\right){\rm H}(1,-1,x)
  \nonumber \\
  &+& \left(\frac{751}{36 (x+1)}+\frac{1069}{36 (x+1)^2}-\frac{983}{3
    (x+1)^3}+\frac{4142}{9 (x+1)^4}-\frac{545}{3 (x+1)^5}-\frac{13}{9
    (x+1)^6}\right){\rm H}(1,0,x) \nonumber \\
  &+& \left(\frac{28}{x+1}+\frac{40}{(x+1)^2}-\frac{440}{(x+1)^3}
  +\frac{620}{(x+1)^4}-\frac{248}{(x+1)^5}\right){\rm H}(1,1,x)
  \nonumber \\
  &+& \left(\frac{3433}{576 (x+1)}+\frac{5935}{144 (x+1)^2}-\frac{25709}{288
    (x+1)^3}+\frac{7405}{144 (x+1)^4}+\frac{385}{96 (x+1)^5}-\frac{2897}{288
    (x+1)^6}+\frac{20}{9 (x+1)^7} \right. \nonumber \\ &&
  \left. +\frac{571}{192 (x-1)}+\frac{81}{32
    (x-1)^2}+\frac{567}{16 (x-1)^3}+\frac{405}{8 (x-1)^4}+\frac{81}{4
    (x-1)^5}\right)\zeta_3 \nonumber \\
  &+& \left(\frac{997}{432 (x+1)}-\frac{2839}{216 (x+1)^2}+\frac{4771}{216
    (x+1)^3}-\frac{5221}{432 (x+1)^4}+\frac{5}{4 (x+1)^5}-\frac{5}{12
    (x+1)^6}\right)\pi ^2 {\rm H}(-1,x) \nonumber \\
  &+& \left(-\frac{8375}{27648 (x+1)}+\frac{61711}{6912 (x+1)^2}-\frac{56633}{3456
    (x+1)^3}+\frac{4195}{576 (x+1)^4}+\frac{1625}{864 (x+1)^5}-\frac{269}{216
    (x+1)^6} \right. \nonumber \\ && \left. +\frac{5}{27
    (x+1)^7}+\frac{1693}{9216 (x-1)}+\frac{81}{512
    (x-1)^2}+\frac{567}{256 (x-1)^3}+\frac{405}{128 (x-1)^4}+\frac{81}{64
    (x-1)^5}\right)\pi ^2 {\rm H}(0,x) \nonumber \\
  &+& \left(-\frac{43}{6 (x+1)}+\frac{449}{18 (x+1)^2}-\frac{212}{9
    (x+1)^3}-\frac{164}{9
    (x+1)^4}+\frac{36}{(x+1)^5}-\frac{12}{(x+1)^6}\right){\rm H}(-1,-1,0,x)
  \nonumber \\
  &+& \left(\frac{545}{36 (x+1)}-\frac{154}{3 (x+1)^2}+\frac{1429}{18
    (x+1)^3}-\frac{2059}{36
    (x+1)^4}+\frac{21}{(x+1)^5}-\frac{7}{(x+1)^6}\right){\rm H}(-1,0,-1,x)
  \nonumber \\
  &+& \left(-\frac{1187}{72 (x+1)}+\frac{2581}{36 (x+1)^2}-\frac{1439}{12
    (x+1)^3}+\frac{2009}{24 (x+1)^4}-\frac{57}{2 (x+1)^5}+\frac{19}{2
    (x+1)^6}\right){\rm H}(-1,0,0,x) \nonumber \\
  &+& \left(-\frac{25}{x+1}+\frac{117}{(x+1)^2}-\frac{184}{(x+1)^3}
  +\frac{92}{(x+1)^4}\right){\rm H}(-1,0,1,x)
  \nonumber \\
  &+& \left(\frac{23447}{2304 (x+1)}+\frac{3713}{576 (x+1)^2}-\frac{11467}{288
    (x+1)^3}+\frac{7213}{144 (x+1)^4}-\frac{1345}{36 (x+1)^5}+\frac{1303}{72
    (x+1)^6}-\frac{40}{9 (x+1)^7} \right. \nonumber \\ &&
  \left. +\frac{1315}{768 (x-1)}+\frac{189}{128
    (x-1)^2}+\frac{1323}{64 (x-1)^3}+\frac{945}{32 (x-1)^4}+\frac{189}{16
    (x-1)^5}\right){\rm H}(0,-1,0,x) \nonumber \\
  &+&
  \left(\frac{8}{x+1}+\frac{24}{(x+1)^2}-\frac{56}{(x+1)^3}+\frac{8}{(x+1)^4}
  +\frac{24}{(x+1)^5}-\frac{8}{(x+1)^6}\right){\rm H}(0,-1,1,x) \nonumber \\
  &+& \left(-\frac{4561}{1152 (x+1)}+\frac{10313}{288 (x+1)^2}-\frac{9313}{144
    (x+1)^3}+\frac{1081}{36 (x+1)^4}+\frac{167}{72 (x+1)^5}-\frac{3}{8
    (x+1)^6}-\frac{55}{128 (x-1)} \right. \nonumber \\ && \left. -\frac{27}{64
    (x-1)^2}-\frac{189}{32
    (x-1)^3}-\frac{135}{16 (x-1)^4}-\frac{27}{8
    (x-1)^5}\right){\rm H}(0,0,-1,x)
  \nonumber \\
  &+& \left(\frac{14875}{4608 (x+1)}-\frac{18817}{384 (x+1)^2}+\frac{54893}{576
    (x+1)^3}-\frac{5851}{96 (x+1)^4}+\frac{241}{16 (x+1)^5}-\frac{427}{72
    (x+1)^6}+\frac{10}{9 (x+1)^7} \right. \nonumber \\ &&
  \left. -\frac{1001}{1536 (x-1)}-\frac{135}{256
    (x-1)^2}-\frac{945}{128 (x-1)^3}-\frac{675}{64 (x-1)^4}-\frac{135}{32
    (x-1)^5}\right){\rm H}(0,0,0,x) \nonumber \\
  &+& \left(\frac{43}{72 (x+1)}-\frac{377}{4 (x+1)^2}+\frac{3085}{18
    (x+1)^3}-\frac{392}{9 (x+1)^4}-\frac{172}{3 (x+1)^5}+\frac{248}{9
    (x+1)^6}-\frac{40}{9 (x+1)^7} \right. \nonumber \\ && \left. -\frac{1}{24
    (x-1)}\right){\rm H}(0,0,1,x) \nonumber \\
  &+&
  \left(\frac{8}{x+1}+\frac{24}{(x+1)^2}-\frac{56}{(x+1)^3}+\frac{8}{(x+1)^4}
  +\frac{24}{(x+1)^5}-\frac{8}{(x+1)^6}\right){\rm H}(0,1,-1,x)
  \nonumber \\
  &+& \left(-\frac{1735}{144 (x+1)}-\frac{289}{8 (x+1)^2}+\frac{3107}{36
    (x+1)^3}-\frac{173}{9 (x+1)^4}-\frac{76}{3 (x+1)^5}+\frac{38}{9
    (x+1)^6}+\frac{20}{9 (x+1)^7} \right. \nonumber \\ && \left. +\frac{1}{48
    (x-1)}\right){\rm H}(0,1,0,x) \nonumber \\
  &+& \left(-\frac{16}{x+1}-\frac{48}{(x+1)^2}+\frac{112}{(x+1)^3}
  -\frac{16}{(x+1)^4}-\frac{48}{(x+1)^5}+\frac{16}{(x+1)^6}\right){\rm H}(0,1,1,x)
  \nonumber \\
  &+& \left(-\frac{1}{12 (x+1)}+\frac{5}{12 (x+1)^3}-\frac{5}{6 (x+1)^4}+\frac{3}{4
    (x+1)^5}-\frac{1}{4 (x+1)^6}\right){\rm F}_1(x) \nonumber \\
  &+& \left(-\frac{5}{108 (x+1)}+\frac{5}{54 (x+1)^2}+\frac{5}{108
    (x+1)^3}-\frac{10}{27 (x+1)^4}+\frac{5}{12 (x+1)^5}-\frac{5}{36
    (x+1)^6}\right){\rm F}_2(x) \nonumber \\
  &+& \left(-\frac{4}{27}+\frac{4}{27 (x+1)}+\frac{16}{9 (x+1)^2}-\frac{112}{27
    (x+1)^3}+\frac{16}{27 (x+1)^4}+\frac{16}{3 (x+1)^5}-\frac{128}{27
    (x+1)^6} \right. \nonumber \\ && \left. +\frac{32}{27
    (x+1)^7}\right)\frac{{\rm F}_3(x)}{\sqrt{x^2+6 x+1}} \nonumber \\
  &+& {\rm F}_4(x) \nonumber\\
%
%
&+& \nl~ \left\{- \frac{x^2 (-2 x+(x+1) {\rm H}(0,x)+2)}{16 (x+1)^5}
\right\}\; .
\end{eqnarray}
}

\begin{figure}[t]
  \begin{center}
    \epsfig{file=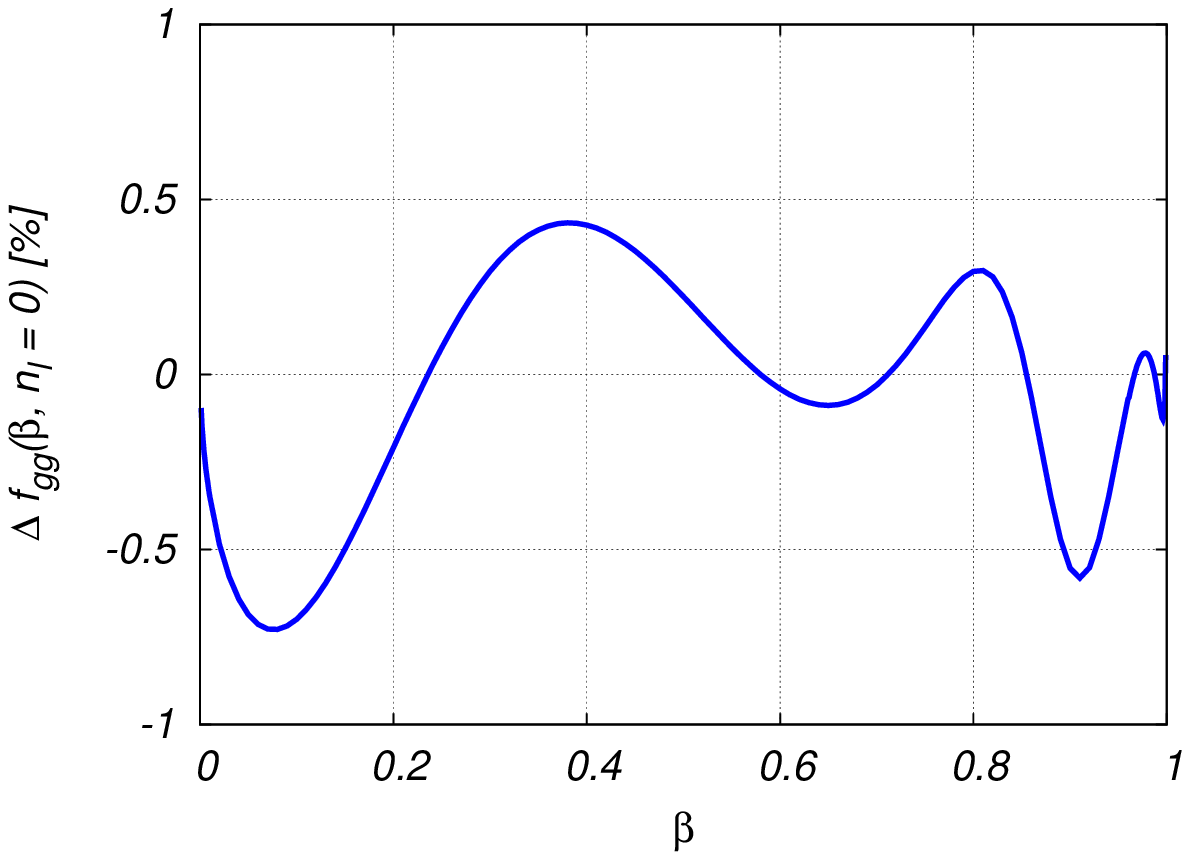,width=8cm}
    \epsfig{file=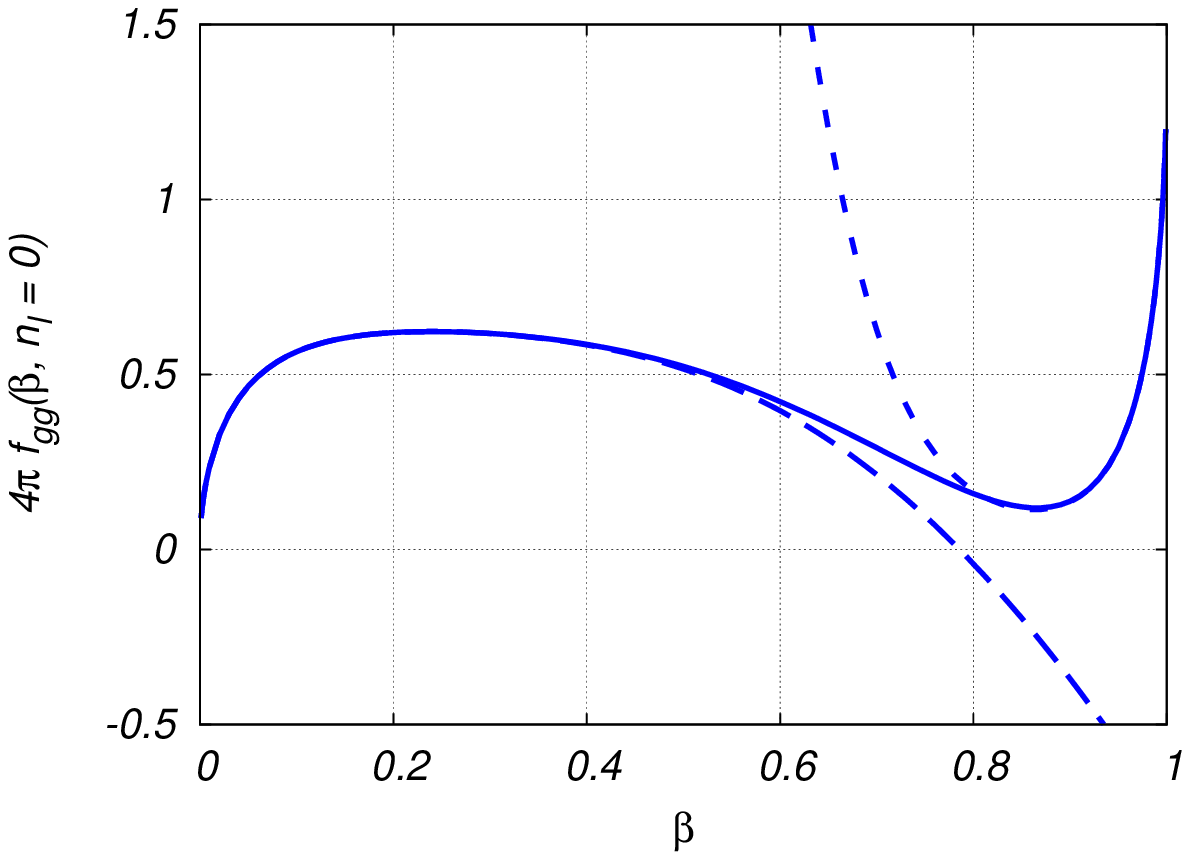,width=8cm}
  \end{center}
  \caption{\it \label{fig:gg} Error implied by the numerical integration of the
    cross section \cite{Nason:1987xz} in the gluon-gluon production channel
    against the exact result. The functions are defined in
    Eqns.~(\ref{eq:fijgg},\ref{eq:error}). The plot
    on the right contains also six terms of the threshold (long-dashed line)
    and the high energy (short-dashed line) expansions.}
\end{figure}

Most of the results above are expressed in terms of the standard
harmonic polylogarithms (HPL) \cite{Remiddi:1999ew} of weight up to
three and of the simple argument $x$, which implies that they can
also be rewritten in terms of the usual polylogarithms of more
complicated arguments but same weight. As we already mentioned in
Section~\ref{sec:partonlevel} and further elaborate in
Section~\ref{sec:singularities}, the presence of additional
singularities beyond those naively expected makes it impossible to
integrate the gluon-gluon cross section in terms of HPL's alone.
Indeed, four additional functions are need. The first three of them
are chosen as follows
\begin{eqnarray}
{\rm F}_1(x) &=& - \int_x^1 {{\rm d} \; z\;\frac{(2 z+1)
    \left({\rm H}(-1,0,z)+{\rm H}(0,-1,z)-{\rm H}(0,0,z)\right)}{2
    \left(z^2+z+1\right)}} \; , \\ \nonumber \\
{\rm F}_2(x) &=& - \int_x^1 {{\rm d} \; z\; \frac{(2 z+3) \left(12
\;
    {\rm H}(-1,0,z)-6 \; {\rm H}(0,0,z)+\pi ^2\right)}{4 \left(z^2+3 z+1\right)}} \; ,
\\ \nonumber \\
{\rm F}_3(x) &=& + \int_x^1 {{\rm d} \; z\; \frac{5 (z-1) \left(12
\;
    {\rm H}(-1,0,z)-6 \; {\rm H}(0,0,z)+\pi ^2\right)}{8 z \sqrt{z^2+6 z+1}}} \; .
\end{eqnarray}
It is clear from these expressions that both ${\rm F}_1$ and ${\rm
F}_2$ can also be expressed through the usual polylogarithms of up
to weight three. Since the resulting expressions are very lengthy
and cumbersome, yet obtainable by automated software, we prefer to
stick to the integral representations which are easier to
manipulate. Unlike ${\rm F}_{1,2}$, however, the presence of the
square root in the integrand of ${\rm F}_3$ makes it impossible to
give a closed form result in terms of standard special functions in
that case.

Besides being able to provide numerical values for the partonic
cross sections at any given point, we would also like to perform
threshold and high energy expansions. As far as HPL's are concerned
this can easily be done with the help of \cite{Maitre:2005uu}. As it
turns out, ${\rm F}_1$, ${\rm F}_2$ and ${\rm F}_3$ can be expanded
with little additional effort by simply integrating the expansions
of the integrands term-by-term. The missing boundaries are trivial
at threshold, since all integrals vanish there by their very
definition. In the high energy limit, the asymptotic behavior
necessary to complete this procedure is
\begin{eqnarray}
{\rm F}_1\left(\frac{m^2}{s}\right) &=& \frac{13 \zeta_3}{24} +
{\cal O}\left(\frac{m^2}{s}\right) \; , \\ \nonumber \\
{\rm F}_2\left(\frac{m^2}{s}\right) &=& \frac{21 \zeta_3}{10} +
{\cal O}\left(\frac{m^2}{s}\right) \; , \\ \nonumber \\
{\rm F}_3\left(\frac{m^2}{s}\right) &=& -\frac{5}{8} \log ^3
\left(\frac{m^2}{s}\right)+\frac{5\pi ^2}{8} \log
\left(\frac{m^2}{s}\right)+\frac{75 \zeta_3}{16} + {\cal
O}\left(\frac{m^2}{s}\right) \; .
\end{eqnarray}
The origin of the last function, ${\rm F}_4$, is discussed at length
in Section~\ref{sec:master}. We choose to write its integral
representation in terms of the original mass variable $\rho=4m^2/s$,
not $x$, (recall that the two are related through
Eq.~(\ref{eq:x-deff})) because this simplifies the arguments of the
special functions. We have
\begin{equation}
{\rm F}_4(x) = \int_{\rho}^{1} {{\rm d} \; \tau ~ I_4\left(
  \rho, \; \tau \right)} \; ,
\end{equation}
where
\begin{eqnarray}
&&  \!\!\!\! \!\!\!\! \!\!\!\! \!\!\!\! \!\!\!\! \!\!\!
  I_4(\rho,\tau) = \frac{45 \rho}{32 \pi \tau}
  \log \left(\frac{1-\sqrt{1-\tau
    }}{1+\sqrt{1-\tau }}\right) \left(
\frac{\left(\left(\rho ^2+1\right) {\rm K}\left(
   \sqrt{-4\rho }\right)-(\rho -1) {\rm E}\left(
   \sqrt{-4\rho }\right)\right) {\rm K}\left(\frac{1}{\sqrt{4
   \tau +1}}\right)}{\sqrt{4 \tau +1}} \right.
\nonumber \\ && \left. \;\;\;\; \;\;\;\;
+\frac{\left(\left(-4 \rho ^2+3 \rho +1\right)
   {\rm E}\left(\frac{1}{\sqrt{4 \rho
   +1}}\right)+\left(3 \rho ^2-3 \rho -2\right)
   {\rm K}\left(\frac{1}{\sqrt{4 \rho +1}}\right)\right)
   {\rm K}\left(\sqrt{-4\tau }\right)}{\sqrt{4 \rho
   +1}}
  \right)
 \; .
\end{eqnarray}
The integrand is expressed through the complete elliptic function of
the first, K, and second, E, kind defined as
\begin{eqnarray}
\label{eq:Kdef} {\rm K}(k) &=& \int_0^1 {\rm d} \; z
\frac{1}{\sqrt{1-z^2}\sqrt{1-k^2 z^2}} \; .
\\ \nonumber \\
\label{eq:Edef} {\rm E}(k) &=& \int_0^1 {\rm d} \; z \frac{\sqrt{1-k^2
z^2}}{\sqrt{1-z^2}} \; ,
\end{eqnarray}

Obtaining the asymptotic behavior of ${\rm F}_4$ in the two
interesting limits is a substantially more complicated task than in
the case of ${\rm F}_1\dots {\rm F}_3$. This is best done with the
help of differential equations. Here we only reproduce the leading
term of the expansions
\begin{equation}
{\rm F}_4(\beta) = -\frac{69}{80} \beta ^5+{\cal O}\left(\beta
^6\right) \; ,
\end{equation}
and
\begin{eqnarray}
{\rm F}_4\left(\frac{m^2}{s}\right) &=& \left(-\frac{15}{32} \log ^3
\left(\frac{m^2}{s}\right)-\frac{45}{32} \log^2
\left(\frac{m^2}{s}\right) +\frac{27 \pi ^2}{32} \log
\left(\frac{m^2}{s}\right)+\frac{81
  \zeta_3}{8}+\frac{27 \pi ^2}{32}\right) \frac{m^2}{s} \nonumber \\
&+& {\cal O}\left(\frac{m^4}{s^2}\right) \; .
\end{eqnarray}

Having given all of the relevant expressions, we would now like to
address the question of the actual precision of the fitting formulae
derived by numerical integration in \cite{Nason:1987xz}. To that
end, we plot the error defined as
\begin{equation}
\label{eq:error} \Delta f^{(1)}_{ij}(\beta,\nl=0) =
\frac{f^{(1)}_{ij}(\beta, \nl=0) -f^{(1), {\rm
NDE}}_{ij}(\beta,\nl=0)}{f^{(1)}_{ij}(\beta,\nl=0)} \; ,
\end{equation}
where $f^{(1), {\rm NDE}}_{ij}$ has been taken from \cite{Nason:1987xz}, as
function of the velocity, $\beta$, over the full range from the threshold
$\beta=0$ to the high energy limit $\beta = 1$. As can be seen in
Figs.~(\ref{fig:qq},\ref{fig:gq},\ref{fig:gg}) the error exceeds the promised
1\% only in the vicinity of a zero of the cross section correction. This is of
course expected, since there is no particular need or phenomenological
interest to find the exact position of the zeros only to have an improved
fitting formula, and the authors of \cite{Nason:1987xz} have not used this
information to obtain their approximations. Overall, as can be concluded from
these plots, the quality of the fits in the case of $f^{(1)}_{q\bar{q}}$ and
$f^{(1)}_{gq}$ is excellent. However, the error almost  approaches the 1\%
bound in the case of $f^{(1)}_{gg}$.  While some loss of precision is
expected given how complicated a function the gluon-gluon cross section is, we
will see in the following that this slightly larger uncertainty is of
relevance to the threshold behavior of the cross section.

The analytic results presented above make it possible to obtain {\it
exact} expansions in both the massless and the threshold limits. As
far as practical applications at colliders are concerned, the
threshold behavior is more important. We give it here with full
account of the color factors and $T_F$ set to its customary value of
$1/2$
\begin{eqnarray}
f^{(1)}_{q\bar{q}}(\beta) &=& \frac{1}{4 \pi^2}f^{(0)}_{q\bar{q}}(\beta) \left(
\left(C_F-\frac{1}{2} C_A\right) \frac{\pi^2}{2\beta} + 2 C_F \log^2
\left(8\beta^2\right) -(8C_F+C_A) \log \left(8\beta^2\right) \right.
\nonumber \\ &+&
  C_F \left(8-\frac{\pi ^2}{3}+3 \log 2-2 \log ^2 2\right)
+ C_A\left(\frac{77}{9}-\frac{\pi ^2}{4}-2 \log 2\right)
\nonumber \\ &+& \left.
 \nl \left(-\frac{5}{9}+\frac{2 \log 2}{3}\right)-\frac{8}{9}
+ {\cal O}\left(\beta\right) \right) \; , \label{eq:asymQQ} \\ \nonumber \\
f^{(1)}_{gg}(\beta) &=& \frac{1}{4 \pi^2}f^{(0)}_{gg}(\beta) \left(\left(
C_F - \frac{\left(N^2-4\right)C_A}{2(N^2-2)} \right)\frac{\pi^2}{2\beta}+2C_A
\log^2\left(8\beta^2\right) -
\frac{(9N^2-20)C_A}{N^2-2}\log\left(8\beta^2\right)\right.
  \nonumber \\ &+& \left.
  C_A \left(\frac{21 N^2-50}{N^2-2}-\frac{\left(17 N^2-40\right)\pi ^2}{24
    \left(N^2-2\right)} +\frac{\left(N^2-4\right) \log 2}{N^2-2}-2 \log^2 2
  \right) \right.
  \nonumber \\ &+& \left.
   C_F \left(-5+\frac{\pi^2}{4}\right) + {\cal O}(\beta) \right) \; .
   \label{eq:asymGG}
\end{eqnarray}
The expressions for the Born level functions $f^{(0)}_{q\bar{q}}$
and $f^{(0)}_{gg}$ can be found in Appendix~\ref{sec:appA}. The
singular term proportional to $1/\beta$ (Coulomb enhancement) as
well as the coefficients of the logarithms $\log (8\beta^2)$
(soft-gluon enhancement) have been known already from
\cite{Nason:1987xz}.

The $\beta$-independent constant terms are important for the
quantitative understanding of heavy-pair production at threshold.
Numerical approximations were first derived in \cite{Nason:1987xz}.
Recently the authors of Ref.~\cite{Hagiwara:2008df} independently
extracted these terms from earlier calculations
\cite{Kuhn:1992qw,Petrelli:1997ge} on quarkonium production at
hadron colliders. The comparison of the results extracted in
\cite{Hagiwara:2008df} with the one derived from the calculation of
\cite{Nason:1987xz} showed that the two results were incompatible
with each other, both for $q{\bar q}$ and $gg$ scattering. The
numerical size of the discrepancy was found to be significantly
larger than the numerical uncertainty inherent in the results of
Ref.~\cite{Nason:1987xz}.

With the help of our calculation this discrepancy is now resolved
and the conclusions are quite intriguing. In the $q{\bar q}$ channel
the discrepancy turned out to be due to a missing contribution in
the quarkonium calculation, see Ref.~\cite{Hagiwara:2008df} for
details. The situation is quite different in the case of the
gluon-gluon channel. In this reaction our calculation agrees with
the analytic result extracted in Ref.~\cite{Hagiwara:2008df}.
Thus, the source of the discrepancy is solely due to an accidental
magnification of the numerical uncertainty of the calculation in
Ref.~\cite{Nason:1987xz}. Indeed, the numerical counterpart of the
$\beta$-independent constant term in Eq.~(\ref{eq:asymGG})
\begin{eqnarray}
&& \!\!\!\! \!\!\!\! \!\!\!\! \!\!\!\! \!\!\!\! \!\!\!
  C_A \left(\frac{21 N^2-50}{N^2-2}-\frac{\left(17 N^2-40\right)\pi ^2}{24
    \left(N^2-2\right)} +\frac{\left(N^2-4\right) \log 2}{N^2-2}-2 \log^2 2
  \right) + C_F \left(-5+\frac{\pi^2}{4}\right) \nonumber \\
&& = \frac{1111}{21}-\frac{283 \pi ^2}{168}+\frac{15 \log 2}{7}-6 \log ^2 2
\simeq 34.88 \; ,
\end{eqnarray}
is according to \cite{Nason:1987xz}
\begin{equation}
\frac{768 \pi}{7} a_0^{gg} \simeq 37.25 \; .
\end{equation}
The difference between the two results is about 7\%, even though the
fitting formula for the cross section from Ref.~\cite{Nason:1987xz}
is more accurate than 1\%, see Fig.~\ref{fig:gg}. This is not entirely
surprising as the following arguments show. Close to threshold the
term proportional to this coefficient in $f^{(1)}_{gg}$ is damped by a factor
of $\beta$, whereas the dominant contribution comes from logarithms of the
velocity and Coulomb enhanced constant terms. Thus its error is less relevant
for the final precision. For larger values of $\beta$ other terms dominate
and their errors conspire to make the result sufficiently precise. An analysis
showing that this mechanism indeed works, i.e. changing the value of
$a_0^{gg}$ by up to 10\% leaves the total cross section within its 1\% bounds,
has been performed in \cite{Hagiwara:2008df}.

This difference represents a powerful example for the need of
analytic results in certain cases. It will be very interesting to
further study how this effect propagates in the soft-gluon resummed
cross section. Work is in progress \cite{CM2}.

Let us finally turn our attention to the high energy limit of the
cross sections. It is well known that the $t$-channel gluon exchange
leads to a constant behavior of the quark-gluon and gluon-gluon
cross sections in this regime. We derive
\begin{eqnarray}
f^{(1)}_{gq}\left(\frac{m^2}{s}\right) &=& \frac{924 N^2-1025}{10800 N^2 \pi }
+ {\cal O}\left(\frac{m^2}{s}\right) \; , \label{eq:asymGQ}
\\ \nonumber \\
f^{(1)}_{gg}\left(\frac{m^2}{s}\right) &=& \frac{2C_A}{C_F}
f^{(1)}_{gq}\left(\frac{m^2}{s}\right) + {\cal O}\left(\frac{m^2}{s}\right) \; .
\end{eqnarray}
The constant in Eq.~(\ref{eq:asymGQ}) is in perfect agreement with the numerical
value obtained in \cite{Nason:1987xz}. The relation between the two cross
sections as given in the second equation above has been explained there as
well.

The quality of deeper expansions around $\beta=0$ and $\beta=1$ is
shown in Fig.~\ref{fig:gg}. With six terms in both cases, most of
the range is covered satisfactorily. In fact, due to large numerical
cancellations in the vicinity of $\beta=0$, it is better to use
the expansion rather than the exact formula. Let us stress that these
cancellations are a problem only in the case of numerical evaluation. Higher
precision in the numerics always allows to retrieve the correct result.

%
%
\section{Singularities of the differential equations}
\label{sec:singularities}

\begin{figure}[t]
  \begin{center}
    \epsfig{file=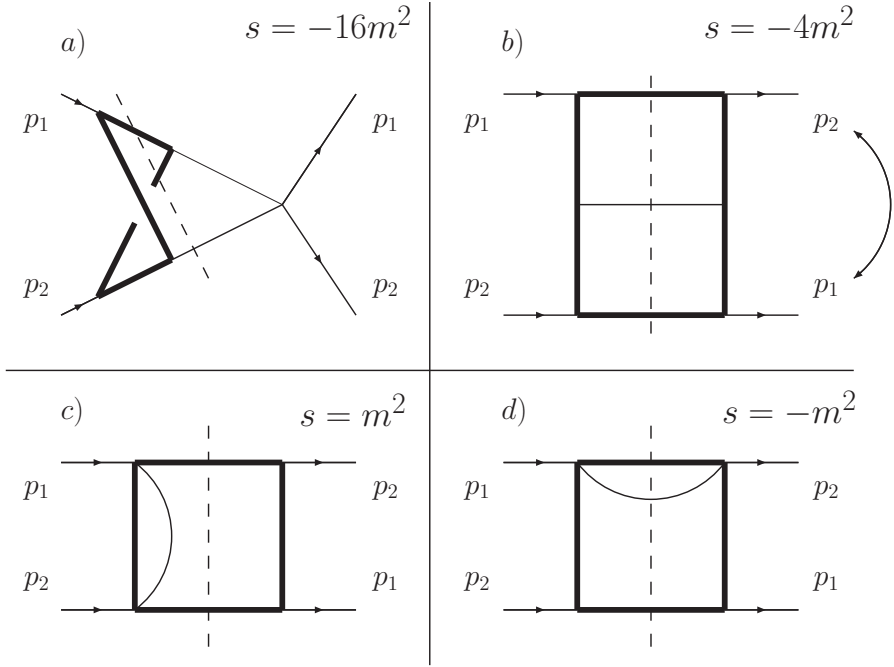,width=12cm}
  \end{center}
  \caption{\it \label{fig:diags} Cut forward Feynman integrals whose differential
    equations introduce a priori unexpected singularities at the listed
    points. Thick lines are massive, whereas thin are massless. Dashed lines
    represent cuts (cut lines are on-shell). Notice the crossed momentum flow
    in the b), c) and d) cases.}
\end{figure}

As already mentioned in Sec.\ref{sec:partonlevel}, besides $s=4m^2$,
$m^2=0$ and $s=0$, we have observed four additional singular points
of the differential equations of the master integrals. These
additional singularities characterize the integrals depicted in
Fig.~\ref{fig:diags}, which contribute exclusively to the
gluon-gluon cross section.

The presence of additional singularities in the differential equations may
have two different implications. One possibility is that, in principle, a
different choice of master integrals may remove these singularities if they do
not have some deeper (physical) meaning. However, if these singularities
persist in the amplitude, it immediately follows that it would not be possible
to express the final result through harmonic polylogarithms of the $x$
variable alone.

We study these alternatives in the following. The main idea behind our
analysis is that, although we work with cut graphs, the homogenous parts of
the corresponding differential equations are the same as for ordinary Feynman
integrals, which means that if the corresponding Feynman integral has a
singularity then the cut graph may have it as well.

It turns out that the singularity at $s=-16m^2$ occurring in the
graph of Fig.~\ref{fig:diags} a) is the one easiest to explain.
Indeed, a quick look at the Landau equations shows that this is the
leading singularity of the Feynman integral, when all lines go on
shell. As such, it can be classified as an anomalous
pseudo-threshold, because it occurs in the Euclidean region.
How to integrate the differential equations in this case is
explained in Section~\ref{sec:master}.

The second singular point, which is also rather easy to explain, is
$s=-4m^2$. Let us take a closer look at Fig.~\ref{fig:diags} b). The graph is
crossed in the forward direction because of the exchanged outgoing
momenta. Its existence is clearly due to the possibility to switch the
identical external gluons. If we treat the graph as a normal box graph with
arbitrary external momenta, then it would have a physical singularity at
$t=4m^2$. The forward limit of the crossed graph corresponds here to the
Mandelstam $u=0$, and not $t=0$. This in turn implies that $t=-s$ and the
usual branch point modeled by a logarithm is mapped to $-s=4m^2$.
Interestingly, this argument further implies that this singularity can occur
in any crossed box graph (in the sense explained above), because the heavy
quark line always has to form a closed loop, and therefore the singular point
$t=4m^2$ is always present.

Interestingly, the seemingly simplest graphs in our problem
have the singularity that is hardest to explain. The singularity of
Fig.~\ref{fig:diags} c) at $s=m^2$ does not correspond to any
leading singularity in the sense of Landau equations. If we kept the
external momenta as free, then there would be three master integrals
for this topology and mass distribution. Their differential
equations do not have more singularities than those of the one-loop
graph without the massless internal line. It is only when we reach
the forward direction that the three integrals collapse to one and
generate the aforementioned singularity in the process. Fortunately,
Fig.~\ref{fig:diags} c) and d) are linked. In fact going from one to
the other while forgetting the cut, can be done by the change $s
\leftrightarrow t$. Since in the forward direction $t=-s$, the
singularity at $s=m^2$ of c) implies a singularity of d) at
$s=-m^2$.

A closer examination of the final result for the $gg$ cross section
Eq.~(\ref{eq:fijgg}) shows that all four singular points described above are
present in the final result in some form or another. Indeed,
Figs.~\ref{fig:diags} a), b), c) and d) correspond to $F_4$, $F_3$, $F_1$ and
$F_2$ respectively. Whereas $s=-16m^2$ for $F_4$ and $s=-m^2$ for $F_2$ are
usual branch points of logarithmic type, $s=-4m^2$ is only a square root type
branch point of $F_3$, and $s=m^2$ is not a branch point of $F_1$ at all. This
is exactly the opposite of what would happen if we had to do with an
amplitude, because there, all points in the Euclidean region would have to be
regular.

As a final remark we would like to stress that while at
next-to-leading order these additional singularities are only
present in the gluon-gluon channel, we anticipate that they will be
present also in the quark-anti-quark channel at the next-to-next-to
leading order.

%
%
\section{The most difficult master integrals}
\label{sec:master}

\begin{figure}[t]
  \begin{center}
    \epsfig{file=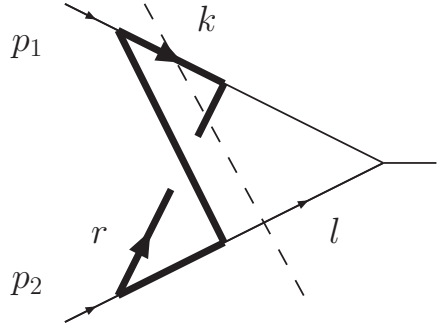,width=6cm}
  \end{center}
  \caption{\it \label{fig:master} The topology and mass distribution of the most
    complicated Feynman integrals, together with the momentum distribution
    necessary to define the numerator of the integrand considered in the
    text. Thin lines are massless, whereas thick have the same non-vanishing
    mass. The dashed line represents a cut (cut lines are on-shell).}
\end{figure}

Of all 37 master integrals occurring in our problem, two stand out
as particularly difficult. They both correspond to the topology and
mass distribution depicted in Fig.~\ref{fig:master} and contribute
to the gluon-gluon cross section exclusively. As usual in such
cases, integration-by-parts (IBP) identities make it possible to
choose any powers of the propagators and irreducible numerators, as
long as one of the integrals cannot be reduced to the other
(independence with respect to IBP's). We make the following choice
\begin{eqnarray}
M_1 &=& s^{6-d} \Gamma^2(1-\epsilon)
\int \frac{{\rm d}^d k \; {\rm d}^d r}{\pi^{d-2}} \;
\frac{1}{\Pi_{i=1}^3 D_i} \delta_+\left(k^2-m^2\right)
\delta_+\left(r^2-m^2\right)\delta_+\left(l^2\right)
\; , \\ \nonumber \\
M_2 &=& s^{4-d} \Gamma^2(1-\epsilon)
\int \frac{{\rm d}^d k \; {\rm d}^d r}{\pi^{d-2}} \;
\frac{(p_1 r)^2}{\Pi_{i=1}^3 D_i}
\delta_+\left(k^2-m^2\right)
\delta_+\left(r^2-m^2\right)\delta_+\left(l^2\right) \; ,
\end{eqnarray}
where $D_i$ are Feynman propagator denominators of the lines which are not
cut, $p_1$ and $r$ are defined in Fig~\ref{fig:master}, whereas $k$ and $l$
are the momenta of the remaining cut lines also shown in
Fig.~\ref{fig:master}. The ``+'' sign of the $\delta$-functions defines the
energy flow of the cut lines, as usual.

One can show that both $M_1$ and $M_2$ are finite (with or without
the cut). Moreover, they enter the cross section with coefficients
regular in the dimensional regularization parameter $\epsilon$,
which means that we only need to know their value at $\epsilon = 0$.
In terms of the dimensionless variable $\rho=4m^2/s$, they satisfy the
following system of differential equations
\begin{eqnarray}
\rho \frac{d M_1}{d \rho} &=& -\frac{2 \rho+1}{4 \rho+1}M_1+\frac{8}{4 \rho+1}M_2
+ H_1 \; , \label{eq:sys} \\ \nonumber \\
\rho \frac{d M_2}{d \rho} &=&-\frac{3 \rho+1}{8 (4 \rho+1)}M_1+\frac{2 \rho+1}{4
\rho+1}M_2 + H_2 \; .
\end{eqnarray}
Although their exact form is irrelevant for our forthcoming discussion we
reproduce here the non-homogenous terms $H_1$ and $H_2$
\begin{eqnarray}
H_1 &=& \frac{5 \left(\log \left(\frac{1-\sqrt{1-\rho}}{1+\sqrt{1-\rho}}\right)+2
  \sqrt{1-\rho}\right)}{4 \rho+1} \; ,\\ \nonumber \\
H_2 &=& \frac{1}{16} H_1 \; .
\end{eqnarray}
Let us rewrite the system as one second order equation for $M_1$. We obtain
\begin{equation}
\label{eq:master} \rho(1+4\rho)M_1''+(1+8\rho)M_1'+M_1 = G \; ,
\end{equation}
where
\begin{equation}
G = (1+4\rho)H_1'+\frac{2\rho-1}{\rho}H_1+\frac{8}{\rho}H_2 \; .
\end{equation}
The homogenous part of Eq.~(\ref{eq:master}) defines a Gauss hypergeometric
function. A change of variable to $k = \sqrt{-4\rho}$, further simplifies the
equation
\begin{equation}
\frac{d}{d k}\left(k\left(1-k^2\right)\frac{d M_1}{dk}\right)-k M_1
= -k G \; .
\end{equation}
The two solutions of the homogenous part are
\begin{eqnarray}
\Phi_1(k) &=& {\rm K}\left(k\right) \; , \\ \label{eq:homo}
\Phi_2(k) &=& {\rm K}\left(\sqrt{1-k^2}\right) \; ,
\end{eqnarray}
where ${\rm K}$ is the complete elliptic function of the first kind
defined in Eq.~(\ref{eq:Kdef}).

As it stands, the solution Eq.~(\ref{eq:homo}) requires analytic
continuation in the physical range, because $k$ is actually
imaginary. Using the relation
\begin{equation}
\Re {\rm K}\left(\sqrt{1-k^2}\right) = \frac{1}{\sqrt{1-k^2}} {\rm
K}\left( \frac{1}{\sqrt{1-k^2}} \right) \; ,
\end{equation}
we choose to set
\begin{equation}
\Phi_2(k) = \frac{1}{\sqrt{1-k^2}} {\rm K}\left(
\frac{1}{\sqrt{1-k^2}} \right) \; .
\end{equation}
At this point, we only need to compute the Wronskian of the
solutions in order to be able to present a complete solution to our
problem. Using the fact that the derivative of the complete elliptic
function of the first kind is
\begin{equation}
\frac{d {\rm K}}{d k} = \frac{{\rm E}(k)}{k(1-k^2)}-\frac{{\rm
K}(k)}{k} \; ,
\end{equation}
where E is the complete elliptic function of the second kind defined in
Eq.~(\ref{eq:Edef}), we obtain the following Wronskian
\begin{equation}
W = \Phi_1 \Phi_2'-\Phi_2 \Phi_1' = -\frac{\pi}{2k(1-k^2)} \; .
\end{equation}
In the above equation we have also used the Legendre relation
\begin{equation}
{\rm E}(k){\rm K}\left(\sqrt{1-k^2}\right) + {\rm K}(k){\rm
E}\left(\sqrt{1-k^2}\right) - {\rm K}(k){\rm
K}\left(\sqrt{1-k^2}\right) = \frac{\pi}{2} \; .
\end{equation}
Finally, the solution to Eq.~(\ref{eq:master}) is
\begin{eqnarray}
M_1(\rho) &=&  C_1 {\rm K}(\sqrt{-4 \rho}) + C_2 \frac{{\rm
K}\left(\frac{1}{\sqrt{4
        \rho+1}}\right)}{\sqrt{4 \rho+1}} \nonumber \\ &-& \frac{1}{\pi}
\int_{\rho}^{1} {\rm d} \; \tau \;
  \left( \frac{{\rm K}\left(\frac{1}{\sqrt{4 \rho+1}}\right) {\rm K}(\sqrt{-4
      \tau})}{\sqrt{4 \rho+1}} - \frac{{\rm K}(\sqrt{-4 \rho})
    {\rm K}\left(\frac{1}{\sqrt{4 \tau+1}}\right)}{\sqrt{4 \tau+1}} \right)
  \nonumber \\ && \;\;\;\; \;\;\;\; \;\;\;\; \;\;\;\; \times
  \tau (4 \tau+1) G(\tau) \; . \label{eq:masRes}
\end{eqnarray}
The choice of the upper bound above allows to easily find the
integration constants $C_1$ and $C_2$. After obtaining $M_2$ from
Eq.~(\ref{eq:sys}), and using the fact that both integrals have to
vanish at threshold, it turns out that $C_{1,2} = 0$.

Before we turn to some general conclusions based on the result we
just obtained, let us stress that in this case the right choice of
integration variable has proven to be invaluable. Indeed, had we
tried to solve the differential equations in our conformal variable
$x$, it would have been impossible to identify the differential
equations as hypergeometric.

A quick inspection of Eq.~(\ref{eq:masRes}) shows that it is
impossible to express this integral with polylogarithms or a
generalization thereof (defined as nested integrals over some
elementary functions). This statement is in particular true of
Goncharov's polylogarithms, and is a simple consequence of the
definition of the elliptic integrals. In general, the presence of
functions with such properties in master integrals does not preclude
the possibility for their cancellation in the final cross section.
However, as the results in Section~\ref{sec:results} demonstrate,
these non-polylogarithmic functions persist also in the final
result.

\begin{figure}[t]
  \begin{center}
    \epsfig{file=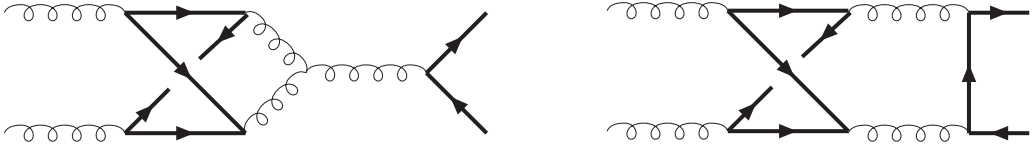,width=14cm}
  \end{center}
  \caption{\it \label{fig:diags2} Example Feynman diagrams contributing to the
    two-loop amplitude for heavy flavor production in gluon fusion, which need
    the integral from Fig.~\ref{fig:master} without the cut.
    Thick lines are massive as before.}
\end{figure}

The fact that the total cross section requires a substantially extended set of
functions is probably not that surprising, because of the never trivial phase
space integration. After all, the two masters considered in this section
correspond to a three particle cut. However, the analysis we have just
performed implies that the same functions will be necessary to solve the
master integrals for the fermionic contributions to the two-loop amplitude for
heavy flavor production in gluon fusion. Examples of Feynman diagrams that
contain the integral from Fig.~\ref{fig:master} without the cut are given in
Fig.~\ref{fig:diags2}.  In this argument it is crucial that the only place
where the presence of a cut plays a role are the non-homogenous terms $H_1$
and $H_2$ (and therefore the function $G$). In consequence, the result for the
two-loop integral will be given by the very same Eq.~(\ref{eq:masRes}). This
means that it will be substantially more difficult to obtain an analytic
result in this case than it was in the quark annihilation channel discussed in
\cite{Bonciani:2008az}. Finally, let us stress that even though elliptic
integrals have been encountered previously in Quantum Field Theory (see for
instance a recent study in \cite{Bailey:2008ib}), we believe that our
observation in this particular problem is an important lesson on complications
associated with massive QCD processes. Whereas in the massless case ordinary
polylogarithms are sufficient, with masses present on internal lines one will
encounter elliptic functions and possibly other yet unexpected objects.

%
%
\section{Summary}
\label{sec:summary}

In this work we have calculated for the first time the exact
analytic expressions for the next-to-leading QCD corrections to all
inclusive cross sections for heavy quark pair production at hadron
colliders. The most important phenomenological application is top
quark pair production at the Tevatron and LHC.

The same corrections have been computed in approximate form by several
groups using more traditional methods based on numerical phase space
integration. Our calculation utilizes completely independent methods
where both real and virtual integrations are performed
simultaneously and in analytic form. We find complete agreement
with the earlier calculations; the comparison to our exact result
shows that the numerical uncertainties in the $q{\bar q}$ and $gq$
reactions are indeed very small. While much larger when compared to the case
of the other two reactions, the uncertainty in the $gg$ reaction does not
exceed the promised accuracy of 1\%.

The motivation for the present calculation was manifold. First, it
is a step in the ambitious program for the calculation of the NNLO
corrections to top quark pair production at the LHC. Indeed, UV and
collinear renormalization require the computation of terms
subleading in the corresponding regulator, or in the case at hand,
beyond $d=4$. Second, one would like to derive and test methods
capable of tackling the very demanding NNLO calculation. Third,
since the resulting complexity is driven by the underlying analytic
structures, the current calculation allows one to gauge the true
level of the complexity of the NNLO calculation. We find for the
first time that new, a priori unexpected analytic structures do
appear even at NLO. What is more important, we are able to pinpoint
their specific origin. Indeed, at NLO we have to compute a total of
37 master integrals and only seven of them have an analytic form
differing from what is known from purely massless calculations.

Therefore, the conclusion that we draw from our present work is that the
analytic methods applied here alow a large degree of automatization and also
seem suitable for the calculation of (at least) the bulk of the NNLO result.

Finally, the analytic results derived in this paper have important
implications to the NLO/NLL (and beyond) phenomenology of top quark pair
production at the Tevatron and LHC, especially for the production close to the
partonic threshold. In particular, we demonstrate that although the
uncertainty in the previous numerical evaluations of the gluon-gluon cross
section is within 1\% of the exact result, the extraction of the so-called
``constant'' terms that are very important at threshold leads to a much larger
numerical difference, one that is in fact as large as 7\%.

Our findings for the threshold behavior of the cross section have to
be carefully explored, especially in light of the fact that the
progress and improvements in the top quark pair production cross
section in the last ten years or so were based on refinements of the
behavior of the cross section at threshold. A detailed investigation
of this effect will be presented elsewhere \cite{CM2}.

%
%
{\bf{Acknowledgments:}}
The research of A.M. is supported by a fellowship from the {\it US
LHC Theory Initiative} through NSF grant {\it PHY-0653342}. Partial
support from the University of Hamburg is also acknowledged. A.M.
would like to thank Daniel Maitre for his kind help with the program
HPL2.0. The work of M.C. was supported in part by the Sofja
Kovalevskaja Award of the   Alexander von Humboldt Foundation   and
by the ToK Program {\it ALGOTOOLS} (MTKD-CD-2004-014319).  Both
authors would like to thank the Galileo Galilei Institute for
Theoretical Physics, Florence, Italy for their hospitality during
the workshop "Advancing Collider Physics: from Twistors to Monte
Carlos" where the current project was initiated. A.M. would also
like to thank G.~Passarino, J.~Smith and G.~Sterman for stimulating
discussions.

\appendix
%
%
\renewcommand{\theequation}{\ref{sec:appA}.\arabic{equation}}
\setcounter{equation}{0}

\section{Some well-known results}
\label{sec:appA}

For convenience of the reader, we collect in this Appendix known
analytic results for different contributions to the partonic cross
sections, which have not been given explicitly in
Section~\ref{sec:results}. We begin by presenting the coefficients
of the scale dependent logarithms. Notice, that because we
renormalize the strong coupling constant with $\nl+1$ active
flavors, these coefficients are not identical to those given in
\cite{Nason:1987xz}. They can be brought into the same form by
decoupling the heavy quark, which at this order of perturbation
theory corresponds to the simple replacement
\begin{equation}
\alpha_s^{(\nl+1)}(\mu^2) = \alpha_s^{(\nl)}(\mu^2) \left( 1 +
\frac{\alpha_s^{(\nl)}(\mu^2)}{6\pi} \log \left( \frac{\mu^2}{m^2}
\right)
 + {\cal O}(\alpha_s^2) \right) \; .
\end{equation}
Clearly, decoupling at next-to-leading order only affects the scale
dependence. This property does not hold at higher orders. In the
case of the $SU(3)$ color group, the $\bar{f}_{ij}$ functions read
\begin{eqnarray}
&& \!\!\!\! \!\!\!\! \!\!\!\! \!\!\!\! \!\!\!\! \!\!\!
  4 \pi \bar{f}^{(1)}_{q\bar{q}}(x) = \nonumber \\ &&
\left(-\frac{484}{243 (x+1)}+\frac{484}{243 (x+1)^2}+\frac{968}{81
  (x+1)^3}-\frac{4840}{243 (x+1)^4}+\frac{1936}{243 (x+1)^5}\right) \nonumber \\
&+&\left(-\frac{32}{27 (x+1)}+\frac{32}{27 (x+1)^2}+\frac{128}{27
  (x+1)^3}-\frac{640}{81 (x+1)^4}+\frac{256}{81 (x+1)^5}\right){\rm H}(0,x)
\nonumber \\
&+&\left(-\frac{128}{81 (x+1)}+\frac{128}{81 (x+1)^2}+\frac{256}{27
  (x+1)^3}-\frac{1280}{81 (x+1)^4}+\frac{512}{81 (x+1)^5}\right){\rm H}(1,x)\nonumber \\
%
%
&+& \nl~\left\{\frac{8 x \left(x^3+3 x^2-3 x-1\right)}{81 (x+1)^5}
\right\} \; ,
\\ \nonumber \\ \nonumber \\
&& \!\!\!\! \!\!\!\! \!\!\!\! \!\!\!\! \!\!\!\! \!\!\!
  4 \pi \bar{f}^{(1)}_{gq}(x) = \nonumber \\ &&
\left(\frac{181}{1620}-\frac{383}{162 (x+1)}+\frac{784}{81
  (x+1)^2}-\frac{1402}{81 (x+1)^3}+\frac{1319}{81 (x+1)^4}-\frac{2638}{405
  (x+1)^5}\right) \nonumber \\
&+&\left(\frac{17}{27 (x+1)}-\frac{61}{54 (x+1)^2}-\frac{1}{27
  (x+1)^3}+\frac{47}{18 (x+1)^4}-\frac{28}{9 (x+1)^5}+\frac{28}{27
  (x+1)^6}\right){\rm H}(0,x) \nonumber \\
&+&\left(-\frac{2}{27 (x+1)}+\frac{2}{9 (x+1)^2}-\frac{8}{27
  (x+1)^3}+\frac{4}{27 (x+1)^4}\right)\pi ^2 \nonumber \\
&+&\left(-\frac{8}{9 (x+1)}+\frac{8}{3 (x+1)^2}-\frac{32}{9
  (x+1)^3}+\frac{16}{9 (x+1)^4}\right){\rm H}(-1,0,x) \nonumber \\
&+&\left(\frac{4}{9 (x+1)}-\frac{4}{3 (x+1)^2}+\frac{16}{9 (x+1)^3}-\frac{8}{9
  (x+1)^4}\right){\rm H}(0,0,x) \; ,
\\ \nonumber \\ \nonumber \\
&& \!\!\!\! \!\!\!\! \!\!\!\! \!\!\!\! \!\!\!\! \!\!\!
  4 \pi \bar{f}^{(1)}_{gg}(x) = \nonumber \\ &&
\left(\frac{181}{360}-\frac{94}{9 (x+1)}+\frac{992}{9 (x+1)^2}-\frac{693}{2
  (x+1)^3}+\frac{14743}{36 (x+1)^4}-\frac{14743}{90 (x+1)^5}\right) \nonumber \\
&+&\left(\frac{173}{18 (x+1)}-\frac{103}{6 (x+1)^2}-\frac{382}{9
  (x+1)^3}+\frac{1081}{9 (x+1)^4}-\frac{269}{3 (x+1)^5}+\frac{176}{9
  (x+1)^6}\right){\rm H}(0,x) \nonumber \\
&+&\left(\frac{7}{x+1}+\frac{10}{(x+1)^2}-\frac{110}{(x+1)^3}+\frac{155}{(x+1)^4}
-\frac{62}{(x+1)^5}\right){\rm H}(1,x)
\nonumber \\
&+&\left(\frac{4}{(x+1)^2}-\frac{23}{3
  (x+1)^3}+\frac{3}{(x+1)^4}+\frac{1}{(x+1)^5}-\frac{1}{3 (x+1)^6}\right)\pi
^2 \nonumber \\
&+&\left(-\frac{8}{x+1}+\frac{24}{(x+1)^2}-\frac{36}{(x+1)^3}+\frac{28}{(x+1)^4}
-\frac{12}{(x+1)^5}+\frac{4}{(x+1)^6}\right){\rm H}(-1,0,x)
\nonumber \\
&+&\left(\frac{2}{x+1}-\frac{18}{(x+1)^2}+\frac{32}{(x+1)^3}
-\frac{16}{(x+1)^4}\right){\rm H}(0,0,x)
\nonumber \\
&+&\left(-\frac{4}{x+1}-\frac{12}{(x+1)^2}+\frac{28}{(x+1)^3}-\frac{4}{(x+1)^4}
-\frac{12}{(x+1)^5}+\frac{4}{(x+1)^6}\right){\rm H}(0,1,x) \; .
\end{eqnarray}
As mentioned in Section~\ref{sec:results}, neither $f_{gq}$ nor
$\bar{f_{gq}}$ depend on $\nl$.

To complete our exposition, we also give the Born contributions,
keeping the full color dependence in concordance with
Eqs.~(\ref{eq:asymQQ}) and (\ref{eq:asymGG}), and setting $T_F$ to
its customary value of $1/2$
\begin{eqnarray}
f^{(0)}_{q\bar{q}}(x) &=& \frac{\pi}{N} C_F \frac{2 x (1-x) \left(x^2+4
  x+1\right)}{3 (x+1)^5} \; , \\ \nonumber \\
f^{(0)}_{gg}(x) &=& \frac{\pi}{N^2-1} \left( C_A \left(\frac{2 (x-1) x
  \left(x^2+7 x+1\right)}{3 (x+1)^5}-\frac{8 x^3
  {\rm H}(0,x)}{(x+1)^6}\right) \right. \\ \nonumber && \left. +C_F \left(\frac{2
  (x-1) x \left(x^2+6 x+1\right)}{(x+1)^5}-\frac{2 x
   \left(x^4+8 x^3+6 x^2+8 x+1\right) {\rm H}(0,x)}{(x+1)^6}\right) \right) \; .
\end{eqnarray}

%
%

\end{document}